\documentclass[11pt,a4paper]{article}

\usepackage{subcaption}
\usepackage{amsmath,mathrsfs}
\usepackage{mathtools}
\usepackage{amssymb}
\usepackage{float}
\usepackage{tikz}
\usepackage{cite}
\usepackage{makecell}
\usetikzlibrary{tikzmark}
\usepackage{jheppub}
\usepackage[title]{appendix}

\begin{document}
\title{Plahte Diagrams for String Scattering Amplitudes}
\author{Pongwit Srisangyingcharoen and Paul Mansfield}

\affiliation{Centre for Particle Theory, University of Durham, Durham DH1 3LE, U.K.}

\emailAdd{pongwit.srisangyingcharoen@durham.ac.uk, p.r.w.mansfield@durham.ac.uk}

\abstract{Plahte identities are monodromy relations between open string scattering amplitudes at tree level derived from the Koba-Nielsen formula. We
represent these
identities by polygons in the complex plane. These diagrams make manifest the appearance of sign changes and singularities in the analytic continuation of amplitudes. They provide a geometric expression of the KLT relations between closed and open string amplitudes.
We also connect the diagrams to the BCFW on-shell recursion relations and generalise them to complex momenta resulting in a relation between the complex phases of partial amplitudes.}


\maketitle
\flushbottom

\section{Introduction}

Kawai, Lewellen and Tyle (KLT) showed that gravity can be seen as the square of gauge theory by discovering how to express closed string amplitudes in terms of products of two open string amplitudes \cite{KLT}. This relation beautifully allows us to compute troublesome gravitational amplitudes by looking at much simpler amplitudes in gauge theory. The connection seems miraclulous as it connects together two theories which are distinct in structure and physical interpretation. This simplification is clear from the  perspective of quantum field theory where the Feynman rules for  gravity contain an infinite number of graviton interaction vertices whilst gauge theory contains only three and four-point interactions.

Understanding string scattering amplitudes has been a fundamental concern among string theorists. An interesting structure was discovered by Plahte in 1970, namely, the Plahte identities which are linear  relations between color-ordered open string scattering amplitudes \cite{Plahte}. The relations are connected to the BCJ relations of  Bern, Carrasco and Johansson  in \cite{BjerrumBohr:2011xe,BjerrumBohr:2010zs}. By using these monodromy relations in string theory, the number of independent color-ordered open string amplitudes with $n$ external legs is reduced from $(n-1)!$ as given by a cyclic property of the trace down to $(n-3)!$ \cite{Minimal,Mixed}. This is in congruence with the field theory of pure Yang-Mills amplitudes where one can use Kleiss-Kujif relations and BCJ relations to represent all color-ordered gauge amplitudes in terms of a basis $(n-3)!$ amplitudes.

The Plahte identities can be illustrated by geometric shapes in the complex plane \cite{Mansfield}. Plahte identities for $n$-point open string amplitudes can be represented by $n$-sided polygons whose sides are proportional to the amplitudes and the angles are given by products of two corresponding momenta. An intriguing result is found in the specific case of the 4-point open tachyon amplitudes where the three amplitudes form a triangle. The area of this triangle is equal to the 4-point closed tachyon amplitude as a consequence of the KLT relations. However, this simple interpretation of the area as a closed string amplitudes is not easy to generalise directly to higher point amplitudes as it is not possible to contain all the relevant open string amplitudes in a single polygon. In this paper we will show what the appropriate generalisation is for five particle scattering and explore further consequences of the Plahte identities. 

\textcolor{black}{
The outline of this paper is as follows. We begin by introducing geometrical diagrams representing the Plahte identities \cite{Plahte} in section two. In section three, we show how the identities can be used to illustrate the  analytic continuation of the Koba Nielsen formula, which requires us to  discuss how these Plahte diagrams are modified when amplitudes become negative and when they diverge. In section four we consider the specific example of the scattering of five particles and formulate the relations between closed and open string amplitudes from the perspective of Plahte diagrams. In section five, we review mixed amplitudes between open and closed strings and present a relation between closed string amplitudes and mixed disk amplitudes for five-particle scattering. A connection between Plahte diagrams and the BCFW recursion relations is described in section six. In section seven, we derive a generalisation of Plahte diagrams to the  meromorphic continuation associated with complex kinematics. We summarise our results in section eight. In the appendices we briefly review the Plahte identities and the KLT relations that are the starting points of our discussion.
}

\section{Plahte Diagrams}
First described by Plahte \cite{Plahte}, Plahte identities are monodromy relations between partial open string scattering amplitudes. An intriguing feature of Plahte identities is that they can be depicted as geometrically. We briefly review the derivation of the Plahte identities in appendix \ref{appen Plahte identities}. 

Let us first explore the simplest example for open string amplitudes, i.e.  the scattering of four tachyons. According to (\ref{Plahte}), the Plahte identity for this process is 
\begin{align}
\mathcal{A}_4(2,1,3,4)+e^{-2\pi i \alpha' k_1\cdot k_2}\mathcal{A}_4(1,2,3,4)+e^{-2\pi i \alpha' k_2\cdot (k_1+k_3)}\mathcal{A}_4(1,3,2,4)=0 \label{Plahte tach}
\end{align}
Combining (\ref{Plahte tach}) with its complex conjugate relation along with the mass-shell condition, $k^2=1/\alpha'$, yields
\begin{equation}
\frac{\mathcal{A}_4(1,2,3,4)}{\sin(2\pi \alpha' k_2\cdot k_4)}=\frac{\mathcal{A}_4(2,1,3,4)}{\sin(2 \pi \alpha' k_2\cdot k_3)}=\frac{\mathcal{A}_4(1,3,2,4)}{\sin(2 \pi \alpha' k_1\cdot k_2)}. \label{sine law}
\end{equation}

\begin{figure}[t]
\centering
\begin{subfigure}{0.48\textwidth}
\resizebox{\linewidth}{!}{
\begin{tikzpicture}
\draw (0,0)--(0:6);
\path (0:6) coordinate (P);
\path (P)++(0:1) coordinate (P1);
\path (P)++(120:3) coordinate (Q);
\path (Q)++(120:1) coordinate (Q1);
\draw (P)--(Q)--(0,0);
\draw (0,0)--(210:1) (P)--(P1) (Q)--(Q1);
\draw (0,0)++(210:.4) arc (210:360:0.4);
\draw (P)++(0:.4) arc (0:120:0.4);
\draw (Q)++(120:.4) arc (120:210:0.4);
\node[below] at (0:3) {$\mathcal{A}_4(2,1,3,4)$};
\node[above left] at (30:2.6) {$\mathcal{A}_4(1,3,2,4)$};
\path (P)++(120:1.5) coordinate (P2);
\node[above right] at (P2) {$\mathcal{A}_4(1,2,3,4)$};
\node[below] at (270:.4) {\small{$-2\pi\alpha'k_2\cdot k_4$}};
\path (P)++(60:.4) coordinate (PP);
\node[above right] at (PP) {\small{$-2\pi\alpha'k_1\cdot k_2$}};
\path (Q)++(165:.4) coordinate (QQ);
\node[above left] at (QQ) {\small{$-2\pi\alpha'k_2\cdot k_3$}};
\end{tikzpicture}
}
\end{subfigure}
\hfill
\begin{subfigure}{0.48\textwidth}
\resizebox{\linewidth}{!}{
\begin{tikzpicture}
\draw (0,0)--(0:6);
\path (0:6) coordinate (P);
\path (P)++(0:1) coordinate (P1);
\path (P)++(120:3) coordinate (Q);
\path (Q)++(120:1) coordinate (Q1);
\draw (P)--(Q)--(0,0);
\draw (0,0)--(210:1) (P)--(P1) (Q)--(Q1);
\draw (0,0)++(210:.4) arc (210:360:0.4);
\draw (P)++(0:.4) arc (0:120:0.4);
\draw (Q)++(120:.4) arc (120:210:0.4);
\node[below] at (0:3) {$\mathcal{A}_4(2,1,3,4)$};
\node[above left] at (30:2.6) {$\mathcal{A}_4(1,3,2,4)$};
\path (P)++(120:1.5) coordinate (P2);
\node[above right] at (P2) {$\mathcal{A}_4(1,2,3,4)$};
\node[below] at (270:.4) {\small{$-2\pi\alpha'k_2\cdot k_4+2\pi$}};
\path (P)++(60:.4) coordinate (PP);
\node[above right] at (PP) {\small{$-2\pi\alpha'k_1\cdot k_2$}};
\path (Q)++(165:.4) coordinate (QQ);
\node[above left] at (QQ) {\small{$-2\pi\alpha'k_2\cdot k_3$}};
\end{tikzpicture}
}
\end{subfigure}
\caption{Triangle representing the 4-point tachyon amplitudes (left) and gauge amplitudes (right) from the Plahte identities.}
\label{triangle}
\end{figure}
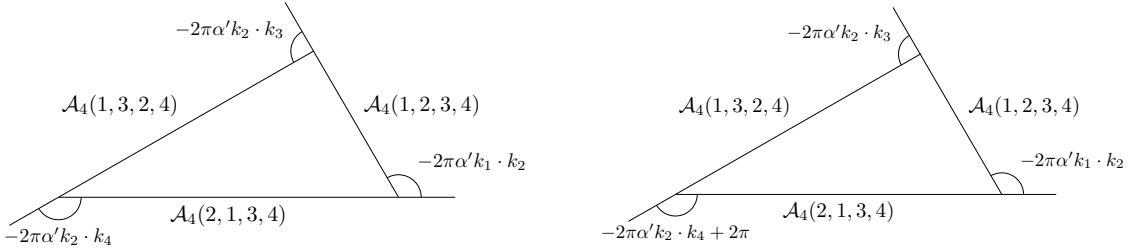

The above relation can be easily pictured as the triangle in fig  \ref{triangle} (left) whose sides refer to the open string amplitudes and angles determined by the product of corresponding momenta. The sum of the external angles is equal to $2\pi$ is guaranteed by the conservation of momentum and the kinematic relation for tachyons \mbox{$k^2=1/ \alpha' $}. Note that the Plahte diagrams are constructed in the kinematic region where all partial amplitudes are positive. The area of this triangle $\Delta$ is quadratic in open string amplitudes and in \cite{Mansfield} the KLT relations were used to show that this is proportional to the four-point tachyonic closed string amplitude $\mathscr{A}_4$  i.e.
\begin{equation}
\mathscr{A}_4=\frac{-i}{4\pi \alpha'} \sin(2\pi \alpha' k_1 \cdot k_2)\mathcal{A}_4(1,2,3,4) \widetilde{\mathcal{A}}_4(2,1,3,4)
=\frac{i}{2\pi \alpha'}\,\Delta
 \label{KLT-4tach}
\end{equation}
where the un-tilded and tilded expressions represent the left-moving and right-moving modes of open string amplitudes respectively. For tachyonic scattering, there is no difference between these modes as \textcolor{black}{there is no involvement of polarization vectors. See appendix \ref{appen KLT} for a brief description of the KLT relations.} 

Unlike the tachyonic case, the Plahte diagrams for gauge particles contain different external angles due to the kinematics, i.e. $k^2=0$. For four gauge particle scattering, the Plahte diagram is again that of figure \ref{triangle} (right). The factor $2\pi$ is added to one of the angles to assure that all angles sum up into full circle. Note that this $2\pi$ phase shift is allowed as it does not change the form of the Plahte identities. The connection between the area of the Plahte diagram and the gauge closed string amplitude is slightly trickier as the area of a diagram for particles with polarisations only refers to the closed string amplitude with polarizations corresponding to those of the open string amplitudes.

According to the figure \ref{triangle} (below), the area of the diagram is
\begin{equation}
\frac{1}{2}\sin(2\pi \alpha' k_1 \cdot k_2)
\xi_{\mu_1}\ldots\xi_{\mu_4}\cdot \xi_{\nu_1}\ldots\xi_{\nu_4}
\mathcal{A}_4^{\mu_1\ldots\mu_4}(1,2,3,4) \mathcal{A}_4^{\nu_1\ldots\nu_4}(2,1,3,4),
\end{equation}
where the open string amplitudes contain the polarization vectors $\xi_i$. One can see that this area corresponds to the gauge closed string amplitude $\mathscr{A}_4=\xi_{\mu_1\nu_1}\ldots\xi_{\mu_4\nu_4}\mathscr{A}_4^{\mu_1\nu_1\ldots\mu_4\nu_4}$ only when the corresponding polarization vector is $\xi_{\mu\nu}=\xi_\mu \xi_\nu$. However, in general we can regain the KLT relation for four-point gauge amplitudes by considering the tensor component of the equation as 
\begin{equation}
\mathscr{A}_4^{\mu_1\nu_1\ldots\mu_4\nu_4}=\frac{-i}{4\pi \alpha'} \sin(2\pi \alpha' k_1 \cdot k_2)
\mathcal{A}_4^{\mu_1\ldots\mu_4}(1,2,3,4) \mathcal{A}_4^{\nu_1\ldots\nu_4}(2,1,3,4).
\end{equation}
The above relation is independent of any polarization vectors which means we can always contract the relation with any polarization vectors we want to consider.
\begin{figure}[t]
\centering
\begin{subfigure}{0.6\textwidth}
\resizebox{\linewidth}{!}{
\begin{tikzpicture}
\path (0,0) coordinate (A);
\path (0:4) coordinate (B);
\path (B)++(72:4) coordinate (C);
\path (C)++(144:4) coordinate (D);
\path (D)++(216:4) coordinate (E);
\path (B)++(0:1) coordinate (B1);
\path (C)++(72:1) coordinate (C1);
\path (D)++(144:1) coordinate (D1);
\path (E)++(216:1) coordinate (E1);
\path (A)++(288:1) coordinate (A1);

\draw (A)--(B) node[pos=0.5,above] {$\mathcal{A}_n(1,2,3,4,\ldots,n)$}--(C) node[pos=0.5,sloped,above] {$\mathcal{A}_n(1,3,2,4,\ldots,n)$}--(D) node[pos=0.6,sloped, above] {$\mathcal{A}_n(1,3,4,2,\ldots,n)$} (E)--(A)node[pos=0.5,sloped,above] {$\mathcal{A}_n(2,1,3,4,\ldots,n)$};
\draw (B)--(B1) (C)--(C1) (D)--(D1) (E)--(E1) (A)--(A1);
\path ([shift=(36:1)]E) coordinate (E2);
\path ([shift=(216:1)]D) coordinate (D2);
\draw (E)--(E2);
\draw (D)--(D2);
\path (E2) -- (D2) node[midway] [sloped] {\ldots};
\draw (A)++(-72:0.5) arc (-72:0:0.5);
\draw (B)++(0:0.5) arc (0:72:0.5);
\draw (C)++(72:0.5) arc (72:144:0.5);
\draw (E)++(216:0.5) arc (216:288:0.5);
\node at ([shift=(-32:.5)]A) [below right] {$-2\pi \alpha' k_1\cdot k_2$};
\node at ([shift=(36:.5)]B) [above right] {$-2\pi \alpha' k_2\cdot k_3$};
\node at (C) [right] {$-2\pi \alpha' k_2\cdot k_4$};
\node at (E) [above left] {$-2\pi \alpha' k_2\cdot k_n$};
\end{tikzpicture}
}
\end{subfigure}
\caption{Plahte diagram for $N$-point open tachyon string amplitudes}
\label{N Plahte}
\end{figure}
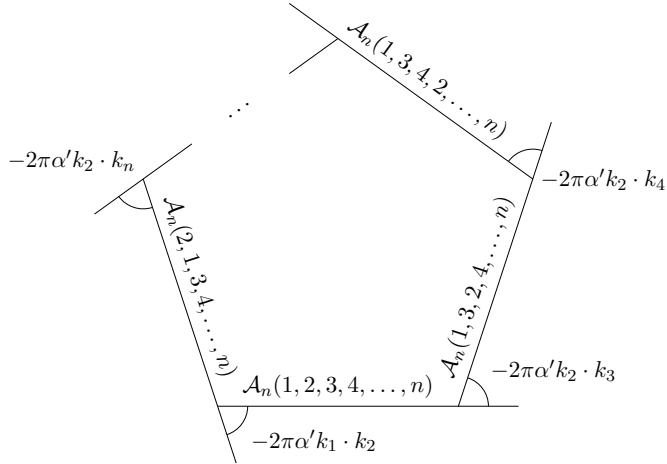

In general, the Plahte identities (\ref{Plahte}) for $n$ particle scattering can be depicted by n-sided polygons in the complex plane whose sides are given by colour-ordered open string amplitudes and its angles correspond to products of two momenta. For tachyon scattering, the Plahte diagram is shown in figure \ref{N Plahte}. To obtain the diagram for gauge particles, one of the external angles need to be added by $2\pi$ as discussed.

However, we need to emphasise that in all the Plahte diagrams we constructed  in this section we took the partial amplitudes to be real, positive and finite. This is not true for  general kinematics as the amplitudes have to be defined by analytic continuation and then it is possible for open string amplitudes to be negative or even divergent. Therefore, we need to take an extra care to construct Plahte diagrams with those features. We will discuss more of these aspects in the next section.

\section{Plahte Diagrams with Negative Amplitudes and Their Dynamics}
So far, in drawing the diagrams in the complex plane, we have taken all amplitudes involved in the diagram to be positive, (as given by the integral expression), but in general this is not the case. The most basic example is the four-point tachyon open string amplitudes. Although the integral expression for the amplitude obtained by Koba and Nielsen \cite{KOBA1969633}, i.e.
\begin{equation}
    \mathcal{A}_4(1,2,3,4)=\int_0^1 dx \ x^{2\alpha' k_1 \cdot k_2} (1-x)^{2 \alpha' k_2 \cdot k_3},
\end{equation}
seems to be positive, it is ill-defined outside the regime where $2\alpha' k_1 \cdot k_2 >-1$ and $2\alpha' k_2 \cdot k_3 >-1$. To obtain the amplitude outside this region, the integral need to be defined by analytic continuation. In this example, it is not hard to see that the expression for this integral is
\begin{equation}
    \mathcal{A}_4(1,2,3,4)=\frac{\Gamma(1+s)\Gamma(1+t)}{\Gamma(2+s+t)} \label{Beta function}
\end{equation}
which provides the analytic continuation to the entire complex plane. We use the notation that $s\equiv 2\alpha'k_1 \cdot k_2$ and $t\equiv 2\alpha'k_2 \cdot k_3$.

\begin{figure}[t]
  \centering
  \begin{subfigure}[t]{0.4\textwidth}
    \centering
    \includegraphics[width=\textwidth]{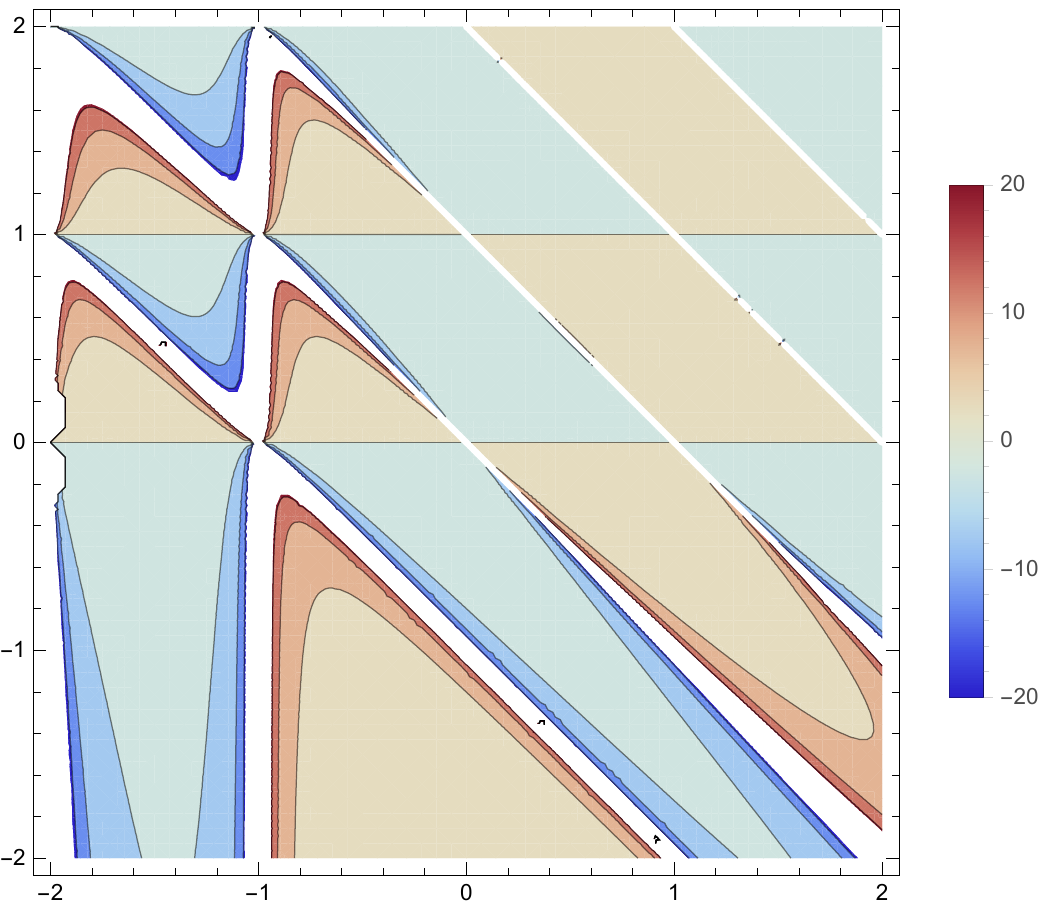}
    \caption*{$\mathcal{A}_4$(2,1,3,4)}
    \vspace{0.1 in}
  \end{subfigure}
\hfill
  \begin{subfigure}[t]{0.4\textwidth}
    \centering
    \includegraphics[width=\textwidth]{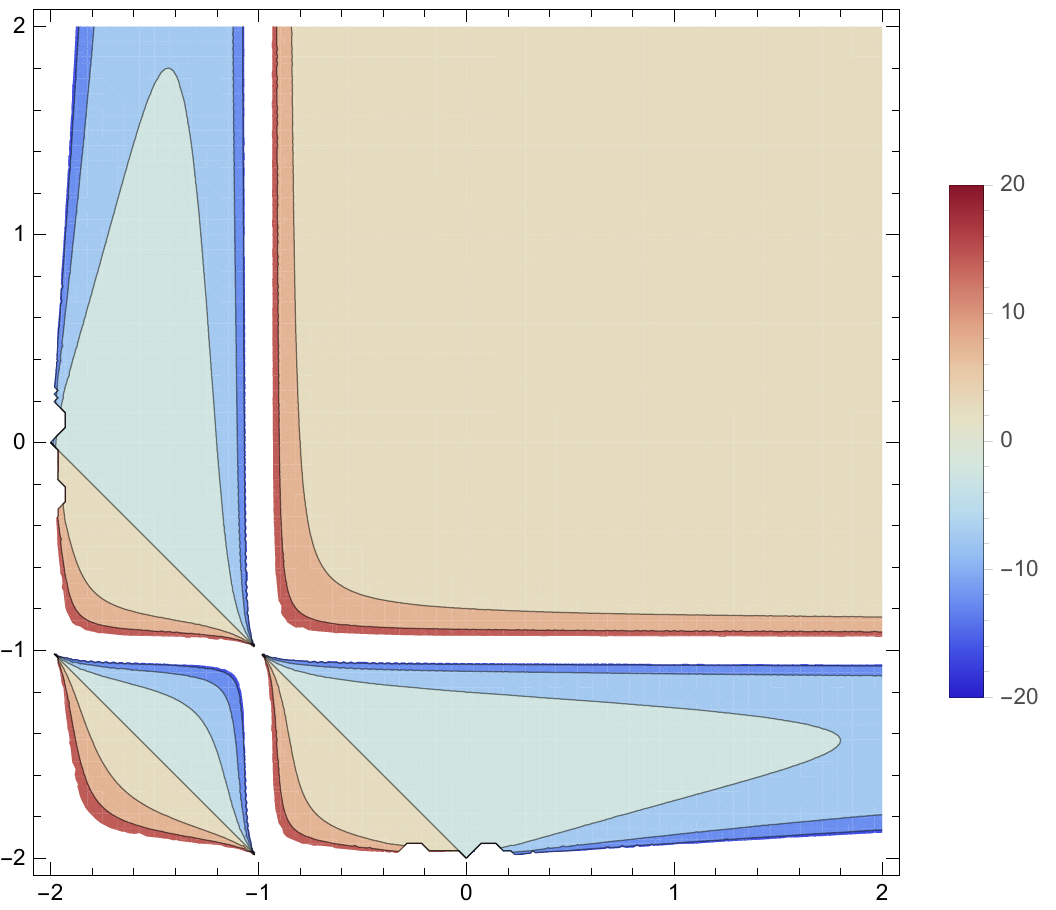}
    \caption*{$\mathcal{A}_4$(1,2,3,4)}
        \vspace{0.1 in}
  \end{subfigure}
\hfill
  \begin{subfigure}[t]{0.4\textwidth}
    \centering
    \includegraphics[width=\textwidth]{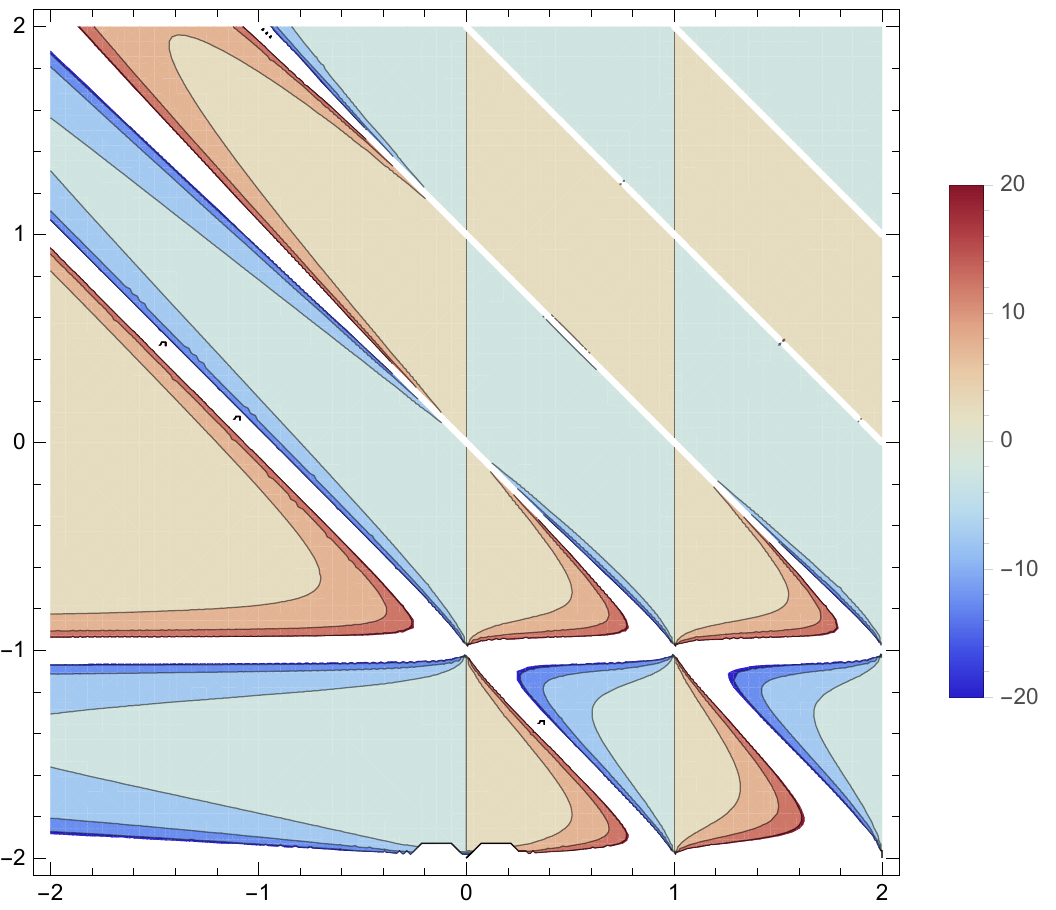}
    \caption*{$\mathcal{A}_4$(1,3,2,4)}
  \end{subfigure}
  \caption{Contour plots for three partial open tachyon amplitudes with $2\alpha'k_1 \cdot k_2$ and $2\alpha'k_2 \cdot k_3$ being X-axis and Y-axis}
  \label{tachyon plot}
\end{figure}

The figure (\ref{tachyon plot}) shows value of all three partial amplitudes for four-point tachyon scattering based on the expression (\ref{Beta function}). The graph was plotted in the kinematic space where $2\alpha'k_1 \cdot k_2$ and $2\alpha'k_2 \cdot k_3$ are X-axis and Y-axis respectively. It is obvious from the figure that some amplitudes become negative. This occurs in regions depicted in blue while the reddish regions show the amplitude being positive. The white areas indicate lines of divergences where the amplitudes are infinitely large.

Alternatively, one can use the Plahte relation in (\ref{sine law}) to provide an analytic continuation of the amplitude beyond the region where the integrals converge. For example, $\mathcal{A}_4(1,2,3,4)$ can be defined with in $t >-1$ and $s+t<-1$ via $\mathcal{A}_4(1,3,2,4)$ using
\begin{equation}
    \mathcal{A}_4(1,2,3,4)=-\frac{\sin{\pi(s+t)}}{\sin{\pi s}}\mathcal{A}_4(1,3,2,4). 
\end{equation}
It is clear from the relation that the amplitude blows up and vanishes when $s$ and $s+t$ is a negative integer respectively. Obviously, it is the same behaviour as would be obtained from the equation (\ref{Beta function}). Similarly, we can evaluate $\mathcal{A}_4(1,2,3,4)$ using $\mathcal{A}_4(1,2,3,4)=-(\sin{\pi(s+t)}/\sin{\pi t})\mathcal{A}_4(2,1,3,4)$ within the region $s >-1$ and $s+t<-1$ in which $\mathcal{A}_4(2,1,3,4)$ is well-defined. 

The remaining region in kinematic space can be determined by considering a map between $\mathcal{A}_4(s-1,t-1)$ and  $\mathcal{A}_4(-s,-t)$ where $\mathcal{A}_4(a,b)$ is defined as the integral $ \int_0^1 dx x^a (1-x)^b$. Consider a product $\mathcal{A}_4(s-1,t-1)\mathcal{A}_4(-s,-t)$
\begin{align}
     &=\int_0^1 dx \int_0^1 dy \ x^{s-1} (1-x)^{t-1} y^{-s} (1-y)^{-t} \nonumber \\
     &= \int_0^\infty dx \int_0^\infty dy \ (x+1)^{-s-t} x^{t-1} (y+1)^{s+t+2} y^{-t}.
\end{align}
A change of variables for $x$ and $y$ as $(x,y)\rightarrow (\frac{1}{x}-1,\frac{1}{y}-1)$ is applied to obtain the last line. The integral can then be performed in plane polar coordinates $(r,\theta)$ along with the change of variable,  $R=(r\cos{\theta}+1)/(r\sin{\theta}+1)$, which yields
\begin{align}
     &\int_0^{\pi/2} d\theta \int_1^{\cot{\theta}} dR \ R^{-s-t} \frac{\cot\theta^t \sec\theta}{(\cos\theta-\sin\theta)} \nonumber \\
     =&\int_0^{\pi/2} d\theta \ \frac{\tan\theta^{s-1}-\tan\theta^{-t}}{(1-s-t)(\cos\theta-\sin\theta)}\sec\theta \nonumber \\
     =&\int_0^\infty dp \frac{p^{s-1}-p^{-t}}{(1-s-t)(1-p)} \label{integral p}
\end{align}
where we have substituted $p=\tan\theta$ in the last line. By splitting the integral into two pieces which are the integral from 0 to 1 and that from 1 to $\infty$, along with a change of variable $p \rightarrow 1/p$ applied to the latter part, the integral takes the form
\begin{equation}
    \frac{1}{1-s-t}\int_0^1 dp \frac{1}{(1-p)} \Bigg[ (p^{s-1}-p^{-s})+(p^{t-1}-p^{-t})  \Bigg] 
\end{equation}

The term $(1-p)^{-1}$ can be Taylor expanded as $\sum_{n=0}^\infty p^n$ allowing the integration to be done to yield the result
\begin{equation}
    \frac{1}{1-s-t} \sum_{n=0}^\infty \Bigg[\Bigg(\frac{1}{s+n}-\frac{1}{n-s+1} \Bigg)+  \Bigg(\frac{1}{t+n}-\frac{1}{n-t+1} \Bigg)\Bigg].
\end{equation}
It can be shown numerically that the above expression is equal to $\pi(\cot(\pi s)+\cot(\pi t))$. Consequently, the relation between $\mathcal{A}_4(s-1,t-1)$ and $\mathcal{A}_4(-s,-t)$ is
\begin{equation}
    \mathcal{A}_4(s-1,t-1)\mathcal{A}_4(-s,-t)=\frac{\pi(\cot(\pi s)+\cot(\pi t))}{1-s-t}. \label{analytic con}
\end{equation}
Note that this is evaluated within $0<s<1$ and $0<t<1$. However, we can still use equation (\ref{analytic con}) as the analytic continuation to determine a value of the amplitude of undetermined points in kinematic space. It is not hard to see that the equation (\ref{analytic con}) can be directly derived from the expression (\ref{Beta function}) as well.

Due to the analytic continuation, it is  clear that the four-point tachyon amplitudes become negative in some regions. The question is how do negative amplitudes affect the Plahte diagram. Since we can always write any amplitude as $\mathcal{A}=(-)\mathcal{A}e^{\pm i \pi}$, the sign of the amplitude can be absorbed by shifting the phase angles next to the amplitude by $+\pi$ or $-\pi$. Each adjacent angle needs to be shifted either by $+\pi$ or $-\pi$ differently in order to keep the sum of the external angles  unchanged at $2\pi$.

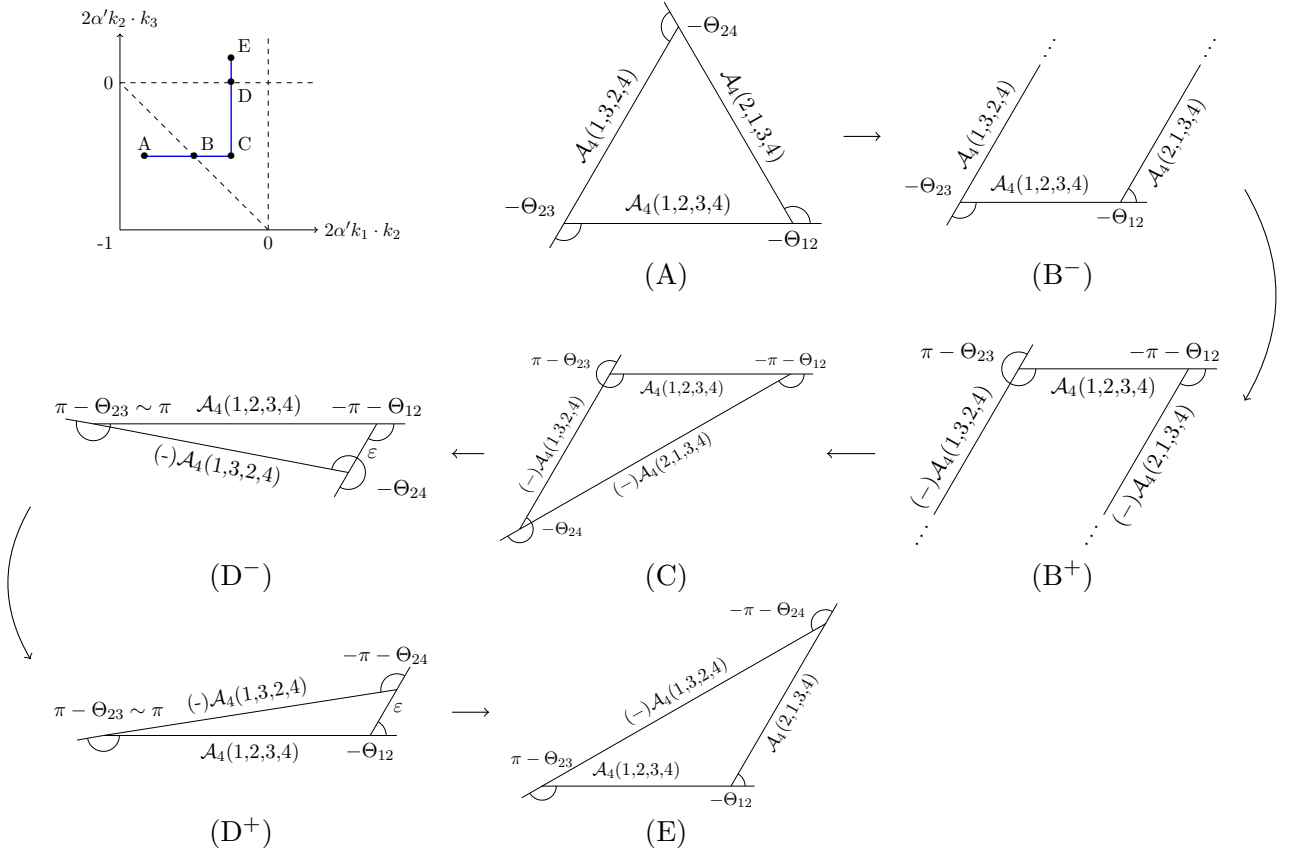
\begin{figure}[hbt!]
\begin{tabular}{ccc}
\begin{subfigure}{.3\textwidth}
  \centering
\resizebox{\linewidth}{!}{
\begin{tikzpicture}
\draw [->] (0,0)--(0,4);
\draw [->](0,0)--(4,0);
\draw [dashed] (0,3)--(4,3);
\draw [dashed] (3,0)--(3,4);
\draw [dashed] (3,0)--(0,3);
\node[below left] at (0,0) {-1};
\node[left] at (0,3) {0};
\node[below] at (3,0) {0};
\node[right] at (4,0) {$2\alpha' k_1 \cdot k_2$};
\node[above] at (0,4) {$2\alpha' k_2 \cdot k_3$};

\draw[blue,thick] (0.5,1.5)--(2.25,1.5) (2.25,1.5)--(2.25,3.5);
\node at (0.5,1.5) {\small{\textbullet}};
\node at (1.5,1.5) {\small{\textbullet}};
\node at (2.25,1.5) {\small{\textbullet}};
\node at (2.25,3) {\small{\textbullet}};
\node at (2.25,3.5) {\small{\textbullet}};

\node[above] at (0.5,1.5) {A};
\node[above right] at (1.5,1.5) {B};
\node[above right] at (2.25,1.5) {C};
\node[below right] at (2.25,3) {D};
\node[above right] at (2.25,3.5) {E};
\end{tikzpicture}
}
\end{subfigure}

&
\begin{subfigure}{.3\textwidth}
  \centering
\resizebox{\linewidth}{!}{
\begin{tikzpicture}
\path (0,0) coordinate (A);
\path (0,0)++(0:4) coordinate (B);
\path (B)++(120:4) coordinate (C);
\draw (A)--(B)node[pos=0.5, sloped, above]{$\mathcal{A}_4$(1,2,3,4)}--(C)node[pos=0.5, sloped, above]{$\mathcal{A}_4$(2,1,3,4)}--(A)node[pos=0.5, sloped, above]{$\mathcal{A}_4$(1,3,2,4)};
\path (B)++(0:0.5) coordinate (BB);
\path (C)++(120:0.5) coordinate (CC);
\path (A)++(-120:0.5) coordinate (AA);
\draw (A)--(AA) (B)--(BB) (C)--(CC);

\draw (A)++(240:0.3) arc (240:360:0.3);
\draw (B)++(0:0.3) arc (0:120:0.3);
\draw (C)++(120:0.3) arc (120:240:0.3);

\node[above left] at (A) {$-\Theta_{23}$};
\node[below] at (B) {$-\Theta_{12}$};
\node[right] at (C) {$-\Theta_{24}$};

\end{tikzpicture}

}
\end{subfigure}
\tikzmark{a}
&
\tikzmark{b}
\begin{subfigure}{.3\textwidth}
  \centering
\resizebox{\linewidth}{!}{
\begin{tikzpicture}
\path (0,0) coordinate (A);
\path (0,0)++(0:3) coordinate (B);
\path (B)++(120:3) coordinate (D);
\path (B)++(60:3) coordinate (C);
\draw (A)--(B)node[pos=0.5, sloped, above]{$\mathcal{A}_4$(1,2,3,4)};
\draw (A)--(D) node[pos=0.5,sloped, above]{$\mathcal{A}_4$(1,3,2,4)} node[pos=1, sloped, right]{\ldots};
\draw (B)--(C) node[pos=0.5,sloped, below]{$\mathcal{A}_4$(2,1,3,4)} node[pos=1, sloped, right]{\ldots};
\path (A)++(240:0.5) coordinate (AA);
\path (B)++(0:0.5) coordinate (BB);
\draw (A)--(AA) (B)--(BB);
\draw (A)++(240:0.3) arc (240:360:0.3);
\draw (B)++(0:0.3) arc (0:60:0.3);
\node[above left] at (A) {$-\Theta_{23}$};
\node[below] at (B) {$-\Theta_{12}$};
\end{tikzpicture}
}
\end{subfigure}

\tikzmark{bb}
\\

& (A) & $(\text{B}^-)$\\

\\
\tikzmark{d}
\begin{subfigure}{.35\textwidth}
  \centering
\resizebox{\linewidth}{!}{
\begin{tikzpicture}
\path (0,0) coordinate (A);
\path (0,0)++(0:5) coordinate (B);
\path (B)++(-120:1) coordinate (C);
\draw (A)--(B)node[pos=0.55, sloped, above]{$\mathcal{A}_4$(1,2,3,4)};
\draw (B)--(C)node[pos=0.6, right]{$\varepsilon$}  (C)--(A)node[pos=0.5,sloped, below]{(-)$\mathcal{A}_4$(1,3,2,4)};
\path (A)++(170:0.5) coordinate (AA);
\path (B)++(0:0.5) coordinate (BB);
\path (C)++(240:0.5) coordinate (CC);

\draw (A)--(AA) (B)--(BB) (C)--(CC);
\draw (A)++(170:0.3) arc (170:360:0.3);
\draw (B)++(-120:0.3) arc (-120:0:0.3);
\draw (C)++(-120:0.3) arc (-120:170:0.3);
\node[above] at ([xshift=10]A) {$\pi-\Theta_{23}\sim \pi$};
\node[above] at ([xshift=0]B) {$-\pi-\Theta_{12}$};
\node[below right] at ([xshift=10]C) {$-\Theta_{24}$};
\end{tikzpicture}
}
\end{subfigure}
\tikzmark{dd}
&
\tikzmark{c}
\begin{subfigure}{.3\textwidth}
  \centering
\resizebox{\linewidth}{!}{
\begin{tikzpicture}
\path (0,0) coordinate (A);
\path (0,0)++(0:4) coordinate (B);
\path (A)++(-120:4) coordinate (C);
\draw (A)--(B)node[pos=0.4, sloped, below]{$\mathcal{A}_4$(1,2,3,4)};
\draw (B)--(C)node[pos=0.5, sloped, below]{$(-)\mathcal{A}_4$(2,1,3,4)}  (C)--(A)node[pos=0.5, sloped, above]{$(-)\mathcal{A}_4$(1,3,2,4)};
\path (A)++(60:0.5) coordinate (AA);
\path (B)++(0:0.5) coordinate (BB);
\path (C)++(-150:0.5) coordinate (CC);

\draw (A)--(AA) (B)--(BB) (C)--(CC);
\draw (A)++(60:0.3) arc (60:360:0.3);
\draw (B)++(-150:0.3) arc (-150:0:0.3);
\draw (C)++(-150:0.3) arc (-150:60:0.3);
\node[above left] at ([xshift=-8]A) {$\pi-\Theta_{23}$};
\node[above] at ([xshift=0]B) {$-\pi-\Theta_{12}$};
\node[right] at ([xshift=10]C) {$-\Theta_{24}$};
\end{tikzpicture}
}
\end{subfigure}
\tikzmark{cc}

&
\tikzmark{bbbb}
\begin{subfigure}{.3\textwidth}
  \centering
\resizebox{\linewidth}{!}{
\begin{tikzpicture}
\path (0,0) coordinate (A);
\path (0,0)++(0:3) coordinate (B);
\path (B)++(-120:3) coordinate (C);
\path (A)++(-120:3) coordinate (D);
\draw (A)--(B)node[pos=0.5, sloped, below]{$\mathcal{A}_4$(1,2,3,4)};
\draw (A)--(D) node[pos=0.6,sloped, above]{$(-)\mathcal{A}_4$(1,3,2,4)} node[pos=1, sloped, left]{\ldots};
\draw (B)--(C) node[pos=0.6,sloped, below]{$(-)\mathcal{A}_4$(2,1,3,4)} node[pos=1, sloped, left]{\ldots};
\path (A)++(60:0.5) coordinate (AA);
\path (B)++(0:0.5) coordinate (BB);
\draw (A)--(AA) (B)--(BB);
\draw (A)++(60:0.3) arc (60:360:0.3);
\draw (B)++(-120:0.3) arc (-120:0:0.3);
\node[above left] at ([xshift=-8]A) {$\pi-\Theta_{23}$};
\node[above] at ([xshift=-7]B) {$-\pi-\Theta_{12}$};
\end{tikzpicture}
}
\end{subfigure}
\tikzmark{bbb}
\\
$(\text{D}^-)$&(C)&$(\text{B}^+)$
\\
\tikzmark{ddd}
\begin{subfigure}{.35\textwidth}
  \centering
\resizebox{\linewidth}{!}{
\begin{tikzpicture}
\path (0,0) coordinate (A);
\path (0,0)++(0:5) coordinate (B);
\path (B)++(60:1) coordinate (C);
\draw (A)--(B)node[pos=0.55, sloped, below]{$\mathcal{A}_4$(1,2,3,4)};
\draw (B)--(C)node[pos=0.6, right]{$\varepsilon$}  (C)--(A)node[pos=0.5,sloped, above]{(-)$\mathcal{A}_4$(1,3,2,4)};
\path (A)++(-170:0.5) coordinate (AA);
\path (B)++(0:0.5) coordinate (BB);
\path (C)++(60:0.5) coordinate (CC);

\draw (A)--(AA) (B)--(BB) (C)--(CC);
\draw (A)++(-170:0.3) arc (-170:0:0.3);
\draw (B)++(0:0.3) arc (0:60:0.3);
\draw (C)++(60:0.3) arc (60:190:0.3);
\node[above] at ([yshift=5,xshift=3]A) {$\pi-\Theta_{23}\sim \pi$};
\node[below] at ([xshift=0]B) {$-\Theta_{12}$};
\node[above] at ([yshift=10,xshift=-6]C) {$-\pi-\Theta_{24}$};
\end{tikzpicture}
}
\end{subfigure}
\tikzmark{dddd}
& 
\tikzmark{e}
\begin{subfigure}{.3\textwidth}
  \centering
\resizebox{\linewidth}{!}{
\begin{tikzpicture}
\path (0,0) coordinate (A);
\path (0,0)++(0:4) coordinate (B);
\path (B)++(60:4) coordinate (C);
\draw (A)--(B)node[pos=0.5, sloped, above]{$\mathcal{A}_4$(1,2,3,4)}--(C)node[pos=0.5, sloped, below]{$\mathcal{A}_4$(2,1,3,4)}--(A)node[pos=0.5, sloped, above]{$(-)\mathcal{A}_4$(1,3,2,4)};
\path (B)++(0:0.5) coordinate (BB);
\path (C)++(60:0.5) coordinate (CC);
\path (A)++(210:0.5) coordinate (AA);
\draw (A)--(AA) (B)--(BB) (C)--(CC);

\draw[-] (A)++(-150:0.3) arc (-150:0:0.3);
\draw[-] (B)++(0:0.3) arc (0:60:0.3);
\draw[-] (C)++(60:0.3) arc (60:210:0.3);

\node[above] at ([yshift=8]A) {$\pi-\Theta_{23}$};
\node[below] at ([xshift=0]B) {$-\Theta_{12}$};
\node[above left] at ([xshift=-8]C) {$-\pi-\Theta_{24}$};

\end{tikzpicture}
}
\end{subfigure}
\\
$(\text{D}^+)$&(E)

\end{tabular}
  \begin{tikzpicture}[overlay, remember picture]
    \draw [->] ({pic cs:a}) -- ({pic cs:b});
    \draw [->] ([yshift=-20,xshift=0]{pic cs:bb}) to [bend left] ([yshift=20,xshift=0]{pic cs:bbb});
\draw [<-] ({pic cs:dd}) -- ([xshift=0]{pic cs:c});
    \draw [<-] ([xshift=-20]{pic cs:bbbb}) -- ([xshift=10]{pic cs:cc});
        \draw [->] ([yshift=-20,xshift=0]{pic cs:d}) to [bend right] ([yshift=20,xshift=0]{pic cs:ddd});
    \draw [->] ({pic cs:dddd}) -- ({pic cs:e});

  \end{tikzpicture}
\caption{Dynamics of Plahte diagram for four tachyon scattering with the kinematic variables flowing from (A) to (E)}
\label{dynamics}
\end{figure}

To put it into a clearer perspective, let us give the example of a four-point tachyonic Plahte diagram. We will investigate how the Plahte diagram behaves as the kinematic variables flow from point A to E along the blue line in the figure (\ref{dynamics})(top left). $\Theta_{ij}$ is a shorthand for $2\pi\alpha' k_i \cdot k_j$. We start our examination at the point A in which all three color-ordered amplitudes are positive. When it approaches the point B, the amplitudes $\mathcal{A}_4(1,3,2,4)$ and $\mathcal{A}_4(2,1,3,4)$ diverge, thus, close to the left of the  point B, the diagram becomes a pair of infinite parallel lines with $\mathcal{A}_4(1,2,3,4)$ as a finite bridge between those lines shown in figure \ref{dynamics}$(\text{B}^-)$

As the kinematic variables flow past the point B, the amplitude $\mathcal{A}_4(1,3,2,4)$ and $\mathcal{A}_4(2,1,3,4)$ become negative. This leads to a shift for $-\Theta_{12}$ and $-\Theta_{23}$ by $-\pi$ and $\pi$ respectively. This is a clear example of how the changing sign of an amplitude ends up shifting the phase angles by $\pm\pi$. Moving to the point C, the Plahte diagram is now a triangle illustrated in figure \ref{dynamics}(C). Remember that the shifts  we made in the angles do not alter the sum of the external angles as can be easily checked.

Moving towards the point D, $\mathcal{A}_4(2,1,3,4)$ becomes smaller and vanishes at the point D. The diagram is now a line with finite length at this point. When it crosses the point D, the amplitudes $\mathcal{A}_4(1,2,3,4)$ and $(-)\mathcal{A}_4(1,3,2,4)$ are flipped with each other shifting the angles $-\pi-\Theta_{12}$ and $-\Theta_{24}$ to $-\Theta_{12}$ and $-\pi-\Theta_{24}$ respectively. This shift reflects the fact that in this region $\mathcal{A}_4(2,1,3,4)$ is negative and to compensate this the adjacent angles need to be shifted by $\pi$ and $-\pi$.

According to this example, the amplitudes change their signs when their values pass through zero or infinity which is similar to what happens at the points D and B respectively in the figure \ref{dynamics}. 
This corresponds to the  Plahte diagram becoming a line with finite length or a pair of infinitely long parallel lines.

\begin{figure}[t]
  \centering
  \begin{subfigure}[t]{0.4\textwidth}
    \centering
    \includegraphics[width=\textwidth]{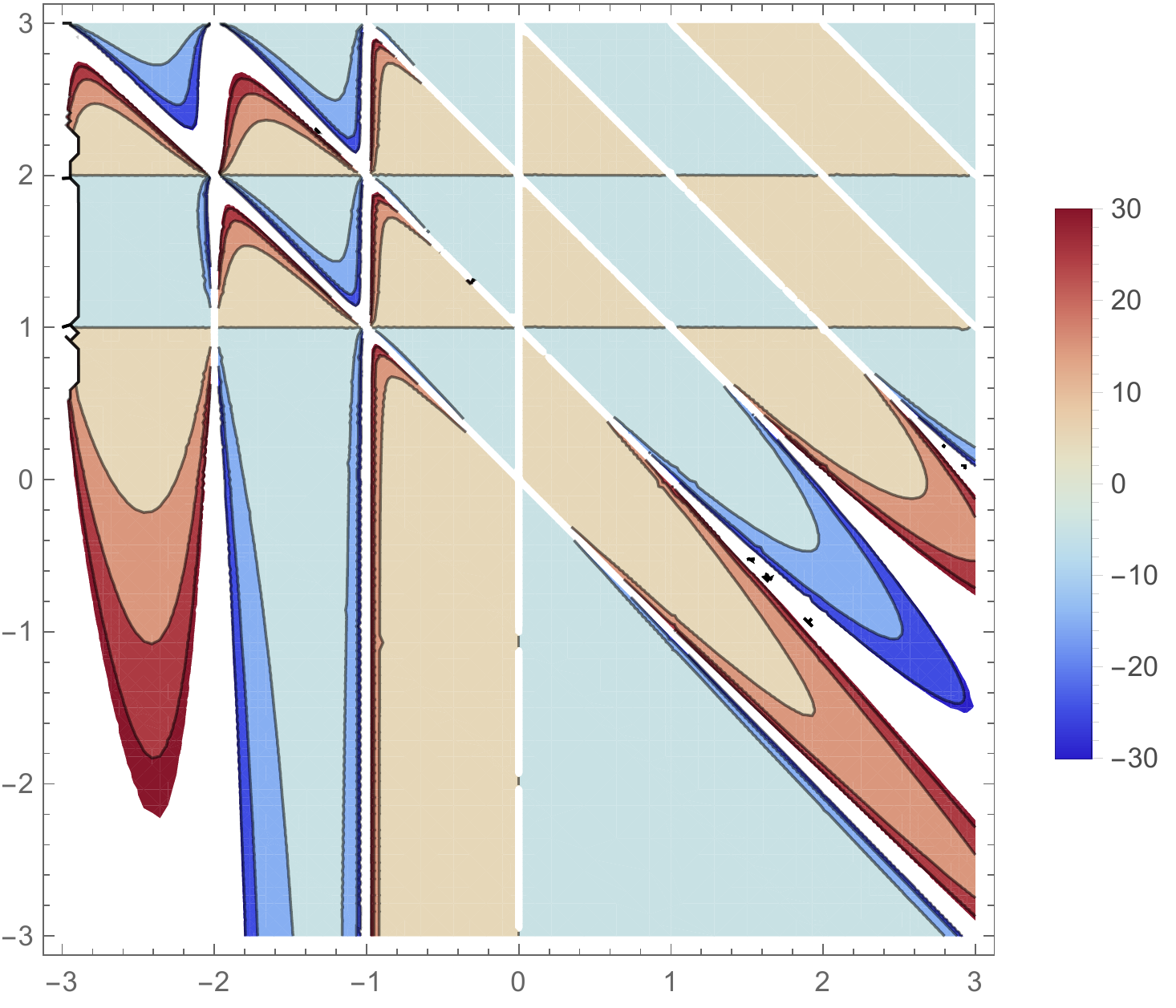}
    \caption*{$\mathcal{A}_4$(2,1,3,4)}
    \vspace{0.1 in}
  \end{subfigure}
\hfill
  \begin{subfigure}[t]{0.4\textwidth}
    \centering
    \includegraphics[width=\textwidth]{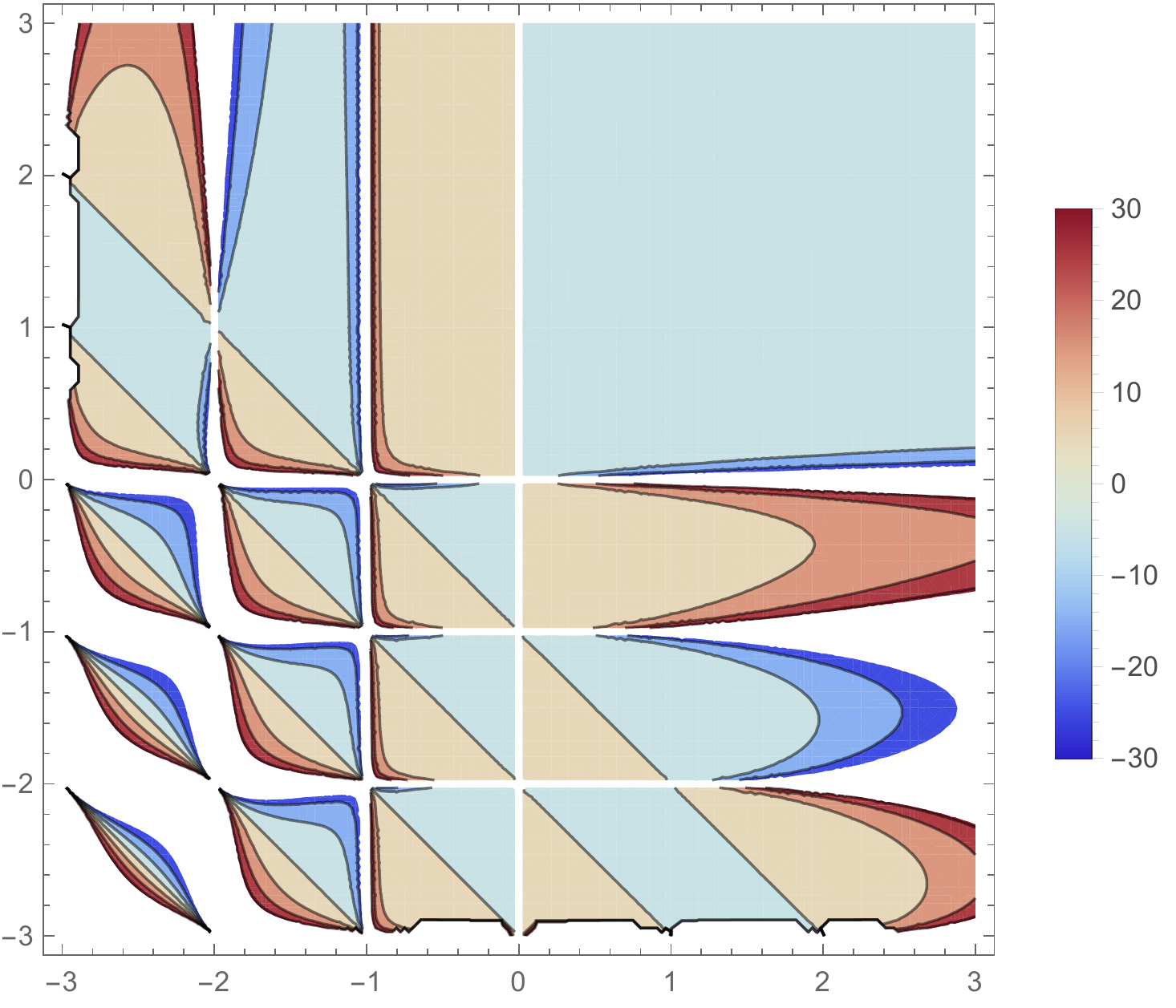}
    \caption*{$\mathcal{A}_4$(1,2,3,4)}
        \vspace{0.1 in}
  \end{subfigure}
\hfill
  \begin{subfigure}[t]{0.4\textwidth}
    \centering
    \includegraphics[width=\textwidth]{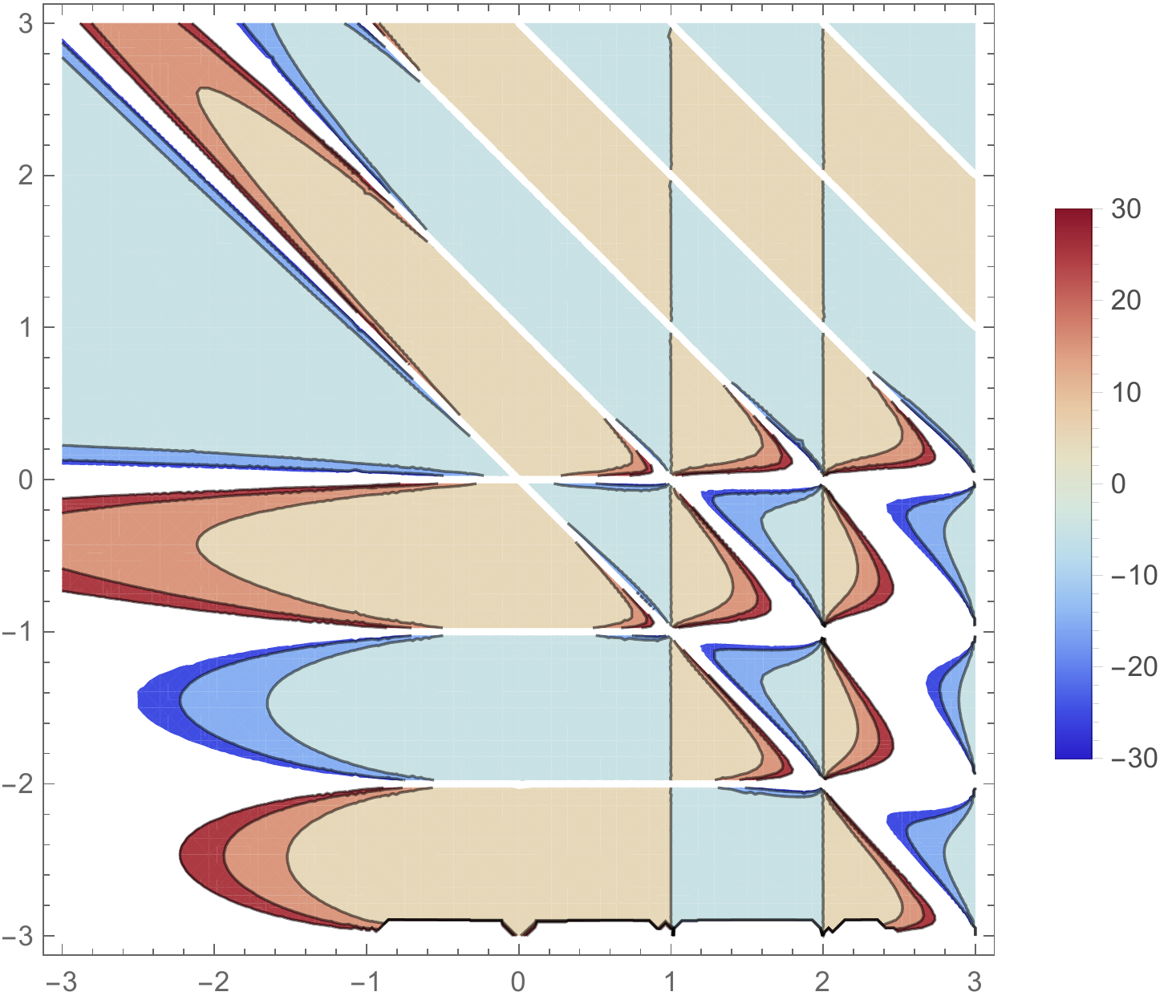}
    \caption*{$\mathcal{A}_4$(1,3,2,4)}
  \end{subfigure}
  \caption{Contour plots for three partial gluon amplitudes with $2\alpha'k_1 \cdot k_2$ and $2\alpha'k_2 \cdot k_3$ being X-axis and Y-axis}
  \label{gluon plot}
\end{figure}

\begin{figure}[thb!]
\begin{tabular}{ccc}
\begin{subfigure}{.3\textwidth}
  \centering
\resizebox{\linewidth}{!}{
\begin{tikzpicture}
\draw [->] (0,0)--(0,4);
\draw [->](0,0)--(4,0);
\draw [dashed] (0,3)--(4,3);
\draw [dashed] (3,0)--(3,4);
\draw [dashed] (3,0)--(0,3);
\node[below left] at (0,0) {-1};
\node[left] at (0,3) {0};
\node[below] at (3,0) {0};
\node[right] at (4,0) {$2\alpha' k_1 \cdot k_2$};
\node[above] at (0,4) {$2\alpha' k_2 \cdot k_3$};

\node at (0.75,1) {A};
\node at (2.25,2) {B};

\end{tikzpicture}
}
\end{subfigure}

&
\begin{subfigure}{.3\textwidth}
  \centering
\resizebox{\linewidth}{!}{
\begin{tikzpicture}
\path (0,0) coordinate (A);
\path (0,0)++(0:4) coordinate (B);
\path (B)++(120:4) coordinate (C);
\draw (A)--(B)node[pos=0.5, sloped, above]{$\mathcal{A}_4$(1,2,3,4)}--(C)node[pos=0.5, sloped, above]{$\mathcal{A}_4$(2,1,3,4)}--(A)node[pos=0.5, sloped, above]{$\mathcal{A}_4$(1,3,2,4)};
\path (B)++(0:0.5) coordinate (BB);
\path (C)++(120:0.5) coordinate (CC);
\path (A)++(-120:0.5) coordinate (AA);
\draw (A)--(AA) (B)--(BB) (C)--(CC);

\draw (A)++(240:0.3) arc (240:360:0.3);
\draw (B)++(0:0.3) arc (0:120:0.3);
\draw (C)++(120:0.3) arc (120:240:0.3);

\node[above left] at (A) {$-\Theta_{23}$};
\node[below] at (B) {$-\Theta_{12}$};
\node[right] at (C) {$-\Theta_{24}+2\pi$};

\end{tikzpicture}

}
\end{subfigure}
&
\begin{subfigure}{.3\textwidth}
  \centering
\resizebox{\linewidth}{!}{
\begin{tikzpicture}
\path (0,0) coordinate (A);
\path (0,0)++(0:4) coordinate (B);
\path (B)++(120:4) coordinate (C);
\draw (A)--(B)node[pos=0.5, sloped, above]{$(-)\mathcal{A}_4$(1,2,3,4)}--(C)node[pos=0.5, sloped, above]{$\mathcal{A}_4$(1,3,2,4)}--(A)node[pos=0.5, sloped, above]{$\mathcal{A}_4$(2,1,3,4)};
\path (B)++(-60:0.5) coordinate (BB);
\path (C)++(60:0.5) coordinate (CC);
\path (A)++(180:0.5) coordinate (AA);
\draw (A)--(AA) (B)--(BB) (C)--(CC);

\draw (A)++(60:0.3) arc (60:180:0.3);
\draw (B)++(-60:0.3) arc (-60:180:0.3);
\draw (C)++(60:0.3) arc (60:300:0.3);

\node[below] at (A) {$-\Theta_{12}-\pi$};
\node[below left] at (B) {$-\Theta_{12}+\pi$};
\node[right] at (C) {$-\Theta_{24}+2\pi$};
\end{tikzpicture}
}
\end{subfigure}
\\
& (A) & (B) 
\end{tabular}
\caption{Plahte diagrams for four gluon scattering in the kinematic regions (A) and (B).}
\label{gluon Plahte}
\end{figure}
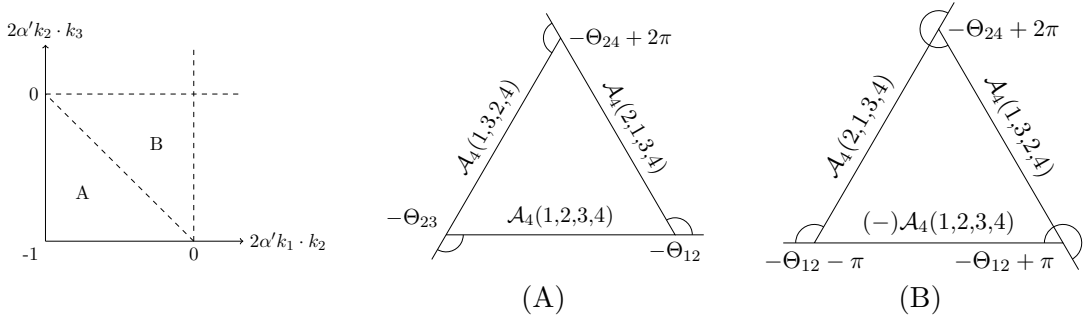

Let us move to another case of interest, Plahte diagrams for gauge bosons. Like the tachyonic case, an open gauge string amplitude can be negative in certain kinematic regions, for instance the region around the origin. This can be seen easily using the BCJ relation,
\begin{equation}
    \frac{A_4(1,2,3,4)}{s_{24}}=\frac{A_4(2,1,3,4)}{s_{23}}=\frac{A_4(1,3,2,4)}{s_{12}},
\end{equation}
where $s_{ij}\equiv 2k_i \cdot k_j$. This is basically the field theory version of (\ref{sine law}). It is unavoidable that at least one partial amplitude must gain a different sign from others as $s_{12}+s_{23}+s_{24}=0$.

The amplitude for multi-gluon scattering can be obtained from superstring theory which directly relates to the Yang-Mills amplitude in the infinite tension limit. The amplitude for four-point gluon scattering is well-known \cite{Green},

\begin{equation}
    \mathcal{A}_4^\text{SUSY}(1,2,3,4)=\frac{\Gamma(1+\alpha's_{12})\Gamma(1+\alpha's_{23})}{\Gamma(1+\alpha's_{12}+\alpha's_{23})}A_4^{\text{YM}}. \label{MHV superstring}
\end{equation}
 
Let consider the Plahte diagram for four gluon scattering assuming that the particle 1 and 2 have negative helicity while the two remaining particles have positive helicity. In this scenario, the 4-point Yang-Mills amplitude $A_4^{\text{YM}}$ can be obtained from the Parke-Taylor formula \cite{Parke-Taylor},
\begin{equation}
    A_4^{\text{YM}}(1^-,2^-,3^+,4^+)=\frac{ \langle 12\rangle^4}{\langle 12\rangle\langle 23\rangle\langle 34\rangle\langle 41\rangle}. \label{Parke-Taylor}
\end{equation}
The above equation is expressed in the spinor-helicity formalism. See \cite{elvang_huang_2015} for more details. We can retrieve the expression for the amplitudes in terms of kinematic momenta by considering the absolute square of the amplitude as
\begin{equation}
     \big|A_4^{\text{YM}}(1^-,2^-,3^+,4^+)\big|^2=\frac{ \langle 12\rangle^4}{\langle 12\rangle\langle 23\rangle\langle 34\rangle\langle 41\rangle} \frac{ [12]^4}{[12][23][34][41]}=\bigg(\frac{s_{12}}{s_{23}}\bigg)^2. \label{Parke-Taylor kinematic}
\end{equation}
Therefore, $A_4^{\text{YM}}(1^-,2^-,3^+,4^+)$ is basically a square root of (\ref{Parke-Taylor kinematic}) up to a certain phase factor. With this calculation, it is not hard to see that all three partial Yang-Mills amplitudes are
\begin{align}
        &A_4^{\text{YM}}(1^-,2^-,3^+,4^+)=\frac{s_{12}}{s_{23}}e^{i\phi_1}, \qquad A_4^{\text{YM}}(2^-,1^-,3^+,4^+)=\frac{s_{12}}{s_{13}}e^{i\phi_2}, \nonumber \\
        \text{and} \quad &A_4^{\text{YM}}(1^-,3^+,2^-,4^+)=\frac{(s_{12})^2}{s_{13}s_{23}}e^{i\phi_3}
\end{align}
where $\phi_i$ is a phase argument corresponding to each amplitude. These phase terms can be determined by BCJ and Kleiss-Kujif relations in equation (\ref{KK}) and (\ref{BCJ}). This allow us to constraint $\phi_1=\phi_2=\phi_3=\phi$. 

Figure \ref{gluon plot} shows the contour plots of all three partial amplitudes for four gluon scattering according to the equation (\ref{MHV superstring}) with $\phi=\pi$. The horizontal and vertical axes of this plot are $2\alpha'k_1 \cdot k_2$ and $2\alpha'k_2 \cdot k_3$ respectively. It is clear from the figure that at one partial amplitude must have a different sign to the others around the origin. Similar to the tachyonic Plahte diagram, the negative amplitudes can be compensated by shifting their adjacent angles by $\pi$ and $-\pi$ differently which can be directly seen in figure \ref{gluon Plahte}.

In figure \ref{gluon Plahte}, the  gluonic Plahte diagrams are constructed for the two different regions depicted in the leftmost figure. All partial amplitudes are positive in region A while the amplitude $\mathcal{A}_4(1,2,3,4)$ is negative in region B. Consequently, the angles next to $\mathcal{A}_4(1,2,3,4)$ are shifted from $-\Theta_{12}$ and $-\Theta_{23}$ to $-\Theta_{12}+\pi$ and $-\Theta_{23}-\pi$ respectively as claimed earlier.

To sum up, in order to draw a Plahte diagram with negative amplitudes, the external angles next to those amplitudes need to be shifted by $\pi$ and $-\pi$ to absorb their negative sign. In general, when a momentum product $2\pi\alpha'k_i \cdot k_j$ is equal to $n\pi$ where $n$ is an integer, it implies a condition where at least one amplitude is about to change its sign. This can be seen by the Plahte identities in (\ref{real Plahte}). 

For four-point scattering, when $2\pi\alpha'k_i \cdot k_j$ is equal to $n\pi$ with an integer $n$, Plahte diagram becomes either a line with finite length or an infinite parallel lines as discussed earlier. The former occurs when the integer $n>l$ while the latter occurs for otherwise. We use the letter $l$ as an identifying integer whose value refers to different types of particles. The values of $l=-1$ and 0 correspond to tachyons and gauge bosons respectively.

\section{Plahte Diagrams for 5-point Amplitudes} \label{section 5-pint}
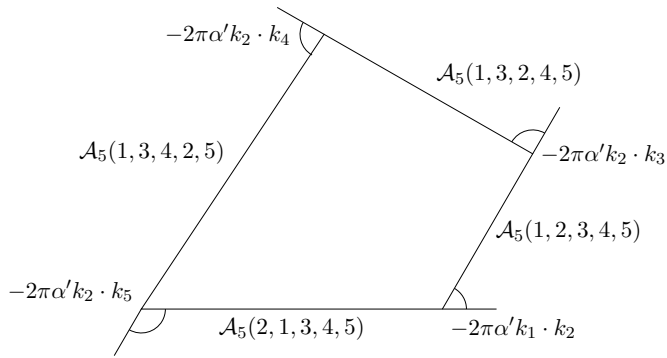
\begin{figure}[t]
\centering
\resizebox{.6\linewidth}{!}{
\begin{tikzpicture}
\path (0:5) coordinate (P);
\path (P)++(60:3) coordinate (Q);
\path (Q)++(150:4) coordinate (R);
\draw (0,0)--(P)--(Q)--(R)--cycle;
\path (-120:.9) coordinate (O1);
\path (P)++(0:.9) coordinate (P1);
\path (Q)++(60:.9) coordinate (Q1);
\path (R)++(150:.9) coordinate (R1);
\draw (0,0)--(O1) (P)--(P1) (Q)--(Q1) (R)--(R1);
\draw ([shift=(-120:0.4)]0,0) arc(-120:0:0.4);
\draw ([shift=(0:0.4)]P) arc(0:60:0.4);
\draw ([shift=(60:0.4)]Q) arc(60:150:0.4);
\draw ([shift=(150:0.4)]R) arc(150:240:0.4);
\node[below] at (2.5,0) {$\footnotesize{\mathcal{A}_5(2,1,3,4,5)}$};
\node[right] at ([shift=(60:1.5)]P) {$\footnotesize{\mathcal{A}_5(1,2,3,4,5)}$};
\node[above right] at ([shift=(150:2)]Q) {$\footnotesize{\mathcal{A}_5(1,3,2,4,5)}$};
\node[above left] at ([shift=(56.5:2.76)]0,0) {$\footnotesize{\mathcal{A}_5(1,3,4,2,5)}$};
\node[above left] at (0,0)
{$-\scriptsize{2\pi\alpha'k_2\cdot k_5}$};
\node[below right] at (P) {$-\scriptsize{2\pi\alpha'k_1\cdot k_2}$};
\node[ right] at (Q) {$-\scriptsize{2\pi\alpha'k_2\cdot k_3}$};
\node[left] at ([xshift=-12]R) {$-\scriptsize{\scriptsize{2\pi\alpha'k_2\cdot k_4}}$};
\end{tikzpicture}
}
\caption{An example of Plahte diagrams for 5-point tachyonic scattering.}
\label{quadrilateral}
\end{figure}

In this section, we will explore the Plahte diagrams for 5-point scattering amplitudes. For simplicity, we will first focus on the scattering of tachyons. The Plahte diagrams for five tachyons is illustrated by quadrilaterals which are directly derived from Plahte identities. Let consider the integral
\begin{align}
\int_{-\infty}^\infty dx_2 \int_0^1 dx_3 \ x_2^{2\alpha' k_1\cdot k_2}x_3^{2\alpha' k_1\cdot k_3} (1-x_2)^{2\alpha' k_2\cdot k_4}(1-x_3)^{2\alpha' k_3\cdot k_4}(x_3-x_2)^{2\alpha' k_2\cdot k_3} \end{align}
where $x_1, x_4$ and $x_5$ are fixed at $x_1=0, x_4=1$ and $x_5=\infty$. This integral corresponds to the Plahte diagram illustrated in figure (\ref{quadrilateral}). The integration  needs to be implemented with careful examination of branch cuts and the use of equation (\ref{z-z}) to provide the correct definition of the partial open string amplitudes.

There are thirty Plahte diagrams in total for five-particle scattering. These can be obtained from similar integrals but with different choices of gauge-fixed and integration variables. The number of diagrams can be halved with the help of reflection symmetry.

Although the Plahte diagrams in figure (\ref{quadrilateral}) are deduced for tachyons, they generalise to other particle states with a suitable phase shift as discussed in the previous section. Polarization vectors are also included in the amplitudes for excited particles. Again, if an amplitude is negative, the diagram needs to be adjusted as discussed previously.

It is unavoidable for Plahte diagrams to share some common sides as there are twelve possible orderings for partial amplitudes and only  fifteen diagrams in total \footnote{Actually they consist of half of all possible ordering as the other amplitudes are related by reflection symmetry.}. Therefore we are able to build a bigger picture by combining diagrams together.

\begin{figure}[h]
\centering
\resizebox{0.7\linewidth}{!}{
\begin{tikzpicture}
\path (135:2.5) coordinate (A);
\path (45:2.5) coordinate (B);
\path (-45:2.5) coordinate (C);
\path (225:2.5) coordinate (D);
\path (135:6) coordinate (E);
\path (45:6) coordinate (F);
\path (-45:6) coordinate (G);
\path (225:6) coordinate (H);
\draw (A)--(B) node[pos=0.5, below] {\footnotesize{$\mathcal{A}_5$(4,2,1,3,5)}} --(C) node[pos=0.5, sloped, below] {\footnotesize{$\mathcal{A}_5$(3,4,2,1,5)}} --(D) node[pos=0.5, above] {\footnotesize{$\mathcal{A}_5$(1,3,4,2,5)}} --(A) node[pos=0.5, sloped, below] {\footnotesize{$\mathcal{A}_5$(2,1,3,4,5)}};
\draw (D)--(E) node[pos=0.5, left] {$\mathcal{A}_5$(3,1,4,2,5)} --(B) node[pos=0.5, sloped, above left] {$\mathcal{A}_5$(3,4,1,2,5)} (A)--(F) node[pos=0.5, sloped, above right] {$\mathcal{A}_5$(1,2,3,4,5)}--(C) node[pos=0.5, right] {$\mathcal{A}_5$(1,3,2,4,5)} (B)--(G) node[pos=0.5,right] {$\mathcal{A}_5$(2,4,1,3,5)}--(D) node[pos=0.5, sloped, below right] {$\mathcal{A}_5$(2,1,4,3,5)} (C)--(H) node[pos=0.5, sloped, below left] {$\mathcal{A}_5$(4,3,2,1,5)}--(A) node[pos=0.5, left] {$\mathcal{A}_5$(
4,2,3,1,5)};

\node[above left] at (A) {A};
\node[above right] at (B) {B};
\node[below right] at (C) {C};
\node[below left] at (D) {D};
\node[above left] at (E) {E};
\node[above right] at (F) {F};
\node[below right] at (G) {G};
\node[below left] at (H) {H};

\end{tikzpicture}
}
\caption{Combined Plahte diagram of 5-point tachyon scattering}
\label{combined}
\end{figure}
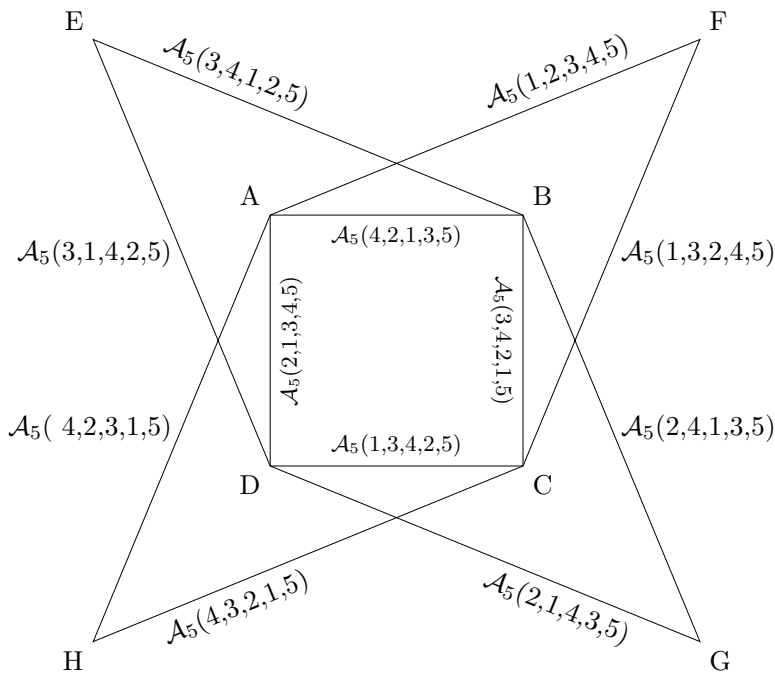

We start by putting the quadrilateral $\square$ABCD in the middle and then attach other quadrilaterals around it. The combined Plahte diagram is shown in figure (\ref{combined}). Obviously, this combined diagram comprises of five different quadrilaterals. Different combined diagrams can be obtained in a similar manner but starting with  a different quadrilateral at the centre. We could extend our combined diagrams still further by considering the peripheral quadrilaterals as new centres. Then we attach another four diagrams surrounding each, however, this would make the resulting diagram quite complicated and hard to analyse. For that reason, we will content ourselves with the diagram as it is presented in figure (\ref{combined}).

Unlike the 4-particle cases, for 5-particles the area of each Plahte diagram is no longer simply proportional to a closed string amplitude. However, this does not mean there is no geometric relation between the closed string amplitudes and the Plahte diagrams. According to the KLT relations for 5-point tachyon string amplitudes,
\begin{align}
\mathscr{A}_5=&\frac{-1}{16\pi^2\alpha'}\mathcal{S}_{k_1,k_2}\mathcal{S}_{k_3,k_4}\mathcal{A}_5(1,2,3,4,5)\mathcal{A}_5(2,1,4,3,5) \nonumber \\&
+ \text{exchange of } (2 \leftrightarrow 3). \label{5KLT}
\end{align}
where $\mathcal{S}_{k_i,k_j}\equiv \sin(2\pi \alpha'k_i\cdot k_j)$. The terms on the right-hand-side of this correspond to areas of parts of 
the combined Plahte diagram of figure(\ref{combined}):
\begin{align}
\mathscr{A}_5=&\frac{-1}{8\pi^2\alpha'} \bigg( \frac{\langle\triangle\text{EBC}\rangle \langle \triangle\text{BCH} \rangle}{(BC)^2}+\frac{\langle\triangle\text{FCD}\rangle  \langle \triangle\text{CDE} \rangle}{(CD)^2}   \nonumber  \\
& +\frac{\langle\triangle\text{GDA}\rangle  \langle \triangle\text{DAF} \rangle}{(AD)^2} +\frac{\langle\triangle\text{HAB}\rangle  \langle \triangle\text{ABG} \rangle}{(AB)^2}\bigg) \label{GeoKLT}
\end{align}
where the angle bracket $\langle \quad \rangle$ denotes the area of the object inside. The elements entering this relation are triangles and squares built on the sides of the central quadrilateral. 
Note that the above argument is generally correct for all types of particles not only for tachyons. Moreover, we can further reshape the equation (\ref{5KLT}) into a new form using elementary Euclidean geometry.

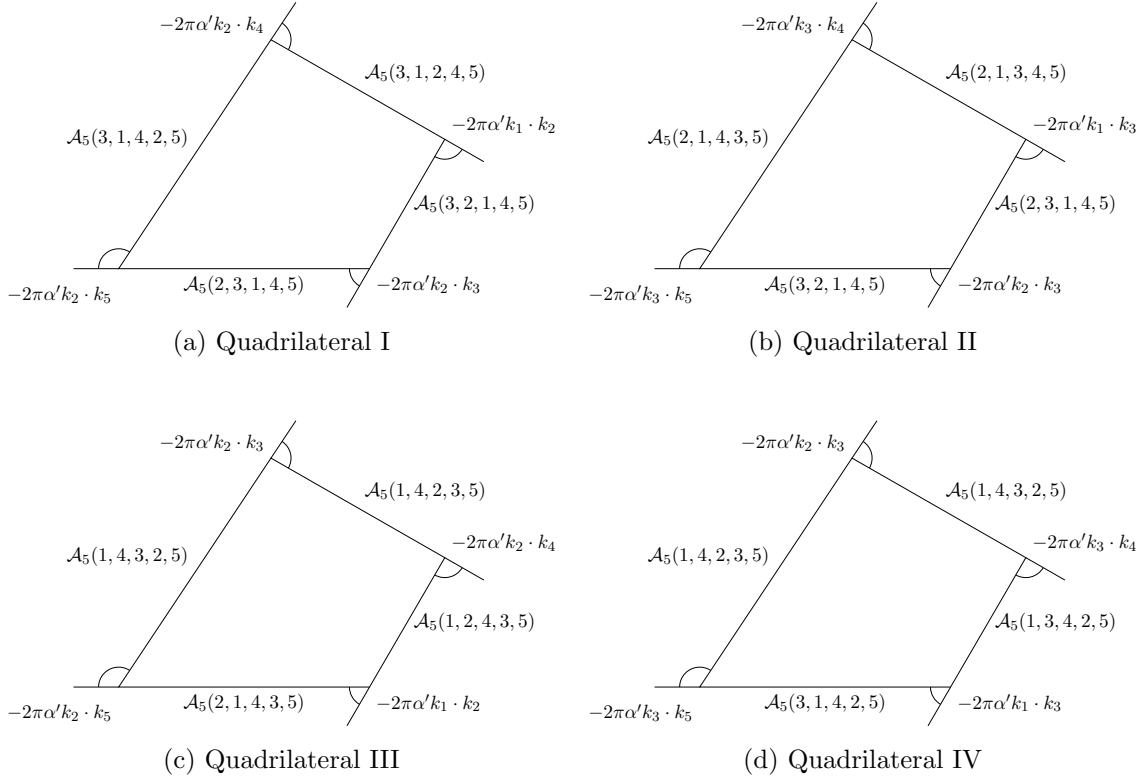
\begin{figure}[t]
\begin{subfigure}[b]{.5\textwidth}
\centering
\resizebox{\linewidth}{!}{
\begin{tikzpicture}
\path (0:5) coordinate (P);
\path (P)++(60:3) coordinate (Q);
\path (Q)++(150:4) coordinate (R);
\draw (0,0)--(P)--(Q)--(R)--cycle;
\path (180:.9) coordinate (O1);
\path (P)++(-120:.9) coordinate (P1);
\path (Q)++(-30:.9) coordinate (Q1);
\path (R)++(56.5:.9) coordinate (R1);
\draw (0,0)--(O1) (P)--(P1) (Q)--(Q1) (R)--(R1);
\draw ([shift=(56.5:0.4)]0,0) arc(56.5:180:0.4);
\draw ([shift=(180:0.4)]P) arc(180:240:0.4);
\draw ([shift=(-30:0.4)]Q) arc(-30:-120:0.4);
\draw ([shift=(56.5:0.4)]R) arc(56.5:-30:0.4);
\node[below] at (2.5,0) {$\footnotesize{\mathcal{A}_5(2,3,1,4,5)}$};
\node[right] at ([shift=(60:1.5)]P) {$\footnotesize{\mathcal{A}_5(3,2,1,4,5)}$};
\node[above right] at ([shift=(150:2)]Q) {$\footnotesize{\mathcal{A}_5(3,1,2,4,5)}$};
\node[above left] at ([shift=(56.5:2.76)]0,0) {$\footnotesize{\mathcal{A}_5(3,1,4,2,5)}$};
\node[below left] at (0,-0.2) {$\scriptsize{-2\pi\alpha'k_2\cdot k_5}$};
\node[below right] at (P) {$\scriptsize{-2\pi\alpha'k_2\cdot k_3}$};
\node[above right] at (Q) {$\scriptsize{-2\pi\alpha'k_1\cdot k_2}$};
\node[above left] at (R) {$\scriptsize{\scriptsize{-2\pi\alpha'k_2\cdot k_4}}$};
\end{tikzpicture}
}
\caption{Quadrilateral I}
\vspace{.3in}
\end{subfigure}
\begin{subfigure}[b]{0.5\textwidth}
\centering
\resizebox{\linewidth}{!}{
\begin{tikzpicture}
\path (0:5) coordinate (P);
\path (P)++(60:3) coordinate (Q);
\path (Q)++(150:4) coordinate (R);
\draw (0,0)--(P)--(Q)--(R)--cycle;
\path (180:.9) coordinate (O1);
\path (P)++(-120:.9) coordinate (P1);
\path (Q)++(-30:.9) coordinate (Q1);
\path (R)++(56.5:.9) coordinate (R1);
\draw (0,0)--(O1) (P)--(P1) (Q)--(Q1) (R)--(R1);
\draw ([shift=(56.5:0.4)]0,0) arc(56.5:180:0.4);
\draw ([shift=(180:0.4)]P) arc(180:240:0.4);
\draw ([shift=(-30:0.4)]Q) arc(-30:-120:0.4);
\draw ([shift=(56.5:0.4)]R) arc(56.5:-30:0.4);
\node[below] at (2.5,0) {$\footnotesize{\mathcal{A}_5(3,2,1,4,5)}$};
\node[right] at ([shift=(60:1.5)]P) {$\footnotesize{\mathcal{A}_5(2,3,1,4,5)}$};
\node[above right] at ([shift=(150:2)]Q) {$\footnotesize{\mathcal{A}_5(2,1,3,4,5)}$};
\node[above left] at ([shift=(56.5:2.76)]0,0) {$\footnotesize{\mathcal{A}_5(2,1,4,3,5)}$};
\node[below left] at (0,-0.2) {$\scriptsize{-2\pi\alpha'k_3\cdot k_5}$};
\node[below right] at (P) {$\scriptsize{-2\pi\alpha'k_2\cdot k_3}$};
\node[above right] at (Q) {$\scriptsize{-2\pi\alpha'k_1\cdot k_3}$};
\node[above left] at (R) {$\scriptsize{\scriptsize{-2\pi\alpha'k_3\cdot k_4}}$};
\end{tikzpicture}
}
\caption{Quadrilateral II}
\vspace{.3in}
\end{subfigure}
\begin{subfigure}[b]{0.5\textwidth}
\centering
\resizebox{\linewidth}{!}{
\begin{tikzpicture}
\path (0:5) coordinate (P);
\path (P)++(60:3) coordinate (Q);
\path (Q)++(150:4) coordinate (R);
\draw (0,0)--(P)--(Q)--(R)--cycle;
\path (180:.9) coordinate (O1);
\path (P)++(-120:.9) coordinate (P1);
\path (Q)++(-30:.9) coordinate (Q1);
\path (R)++(56.5:.9) coordinate (R1);
\draw (0,0)--(O1) (P)--(P1) (Q)--(Q1) (R)--(R1);
\draw ([shift=(56.5:0.4)]0,0) arc(56.5:180:0.4);
\draw ([shift=(180:0.4)]P) arc(180:240:0.4);
\draw ([shift=(-30:0.4)]Q) arc(-30:-120:0.4);
\draw ([shift=(56.5:0.4)]R) arc(56.5:-30:0.4);
\node[below] at (2.5,0) {$\footnotesize{\mathcal{A}_5(2,1,4,3,5)}$};
\node[right] at ([shift=(60:1.5)]P) {$\footnotesize{\mathcal{A}_5(1,2,4,3,5)}$};
\node[above right] at ([shift=(150:2)]Q) {$\footnotesize{\mathcal{A}_5(1,4,2,3,5)}$};
\node[above left] at ([shift=(56.5:2.76)]0,0) {$\footnotesize{\mathcal{A}_5(1,4,3,2,5)}$};
\node[below left] at (0,-0.2)
{$\scriptsize{-2\pi\alpha'k_2\cdot k_5}$};
\node[below right] at (P) {$\scriptsize{-2\pi\alpha'k_1\cdot k_2}$};
\node[above right] at (Q) {$\scriptsize{-2\pi\alpha'k_2\cdot k_4}$};
\node[above left] at (R) {$\scriptsize{\scriptsize{-2\pi\alpha'k_2\cdot k_3}}$};
\end{tikzpicture}
}
\caption{Quadrilateral III}

\end{subfigure}
\begin{subfigure}[b]{0.5\textwidth}
\centering
\resizebox{\linewidth}{!}{
\begin{tikzpicture}
\path (0:5) coordinate (P);
\path (P)++(60:3) coordinate (Q);
\path (Q)++(150:4) coordinate (R);
\draw (0,0)--(P)--(Q)--(R)--cycle;
\path (180:.9) coordinate (O1);
\path (P)++(-120:.9) coordinate (P1);
\path (Q)++(-30:.9) coordinate (Q1);
\path (R)++(56.5:.9) coordinate (R1);
\draw (0,0)--(O1) (P)--(P1) (Q)--(Q1) (R)--(R1);
\draw ([shift=(56.5:0.4)]0,0) arc(56.5:180:0.4);
\draw ([shift=(180:0.4)]P) arc(180:240:0.4);
\draw ([shift=(-30:0.4)]Q) arc(-30:-120:0.4);
\draw ([shift=(56.5:0.4)]R) arc(56.5:-30:0.4);
\node[below] at (2.5,0) {$\footnotesize{\mathcal{A}_5(3,1,4,2,5)}$};
\node[right] at ([shift=(60:1.5)]P) {$\footnotesize{\mathcal{A}_5(1,3,4,2,5)}$};
\node[above right] at ([shift=(150:2)]Q) {$\footnotesize{\mathcal{A}_5(1,4,3,2,5)}$};
\node[above left] at ([shift=(56.5:2.76)]0,0) {$\footnotesize{\mathcal{A}_5(1,4,2,3,5)}$};
\node[below left] at (0,-0.2)
{$\scriptsize{-2\pi\alpha'k_3\cdot k_5}$};
\node[below right] at (P) {$\scriptsize{-2\pi\alpha'k_1\cdot k_3}$};
\node[above right] at (Q) {$\scriptsize{-2\pi\alpha'k_3\cdot k_4}$};
\node[above left] at (R) {$\scriptsize{\scriptsize{-2\pi\alpha'k_2\cdot k_3}}$};
\end{tikzpicture}
}
\caption{Quadrilateral IV}
\end{subfigure}
\caption{Plahte diagrams for 5-point tachyonic scattering}
\label{4quadrilaterals}
\end{figure}

Let us consider the quadrilaterals I and II in figure \ref{4quadrilaterals}. We can extract the Plahte identities for each diagram as 
\begin{subequations}
\begin{align}
\mathcal{S}_{k_2,k_4}\mathcal{A}_5(3,1,4,2,5)=&\mathcal{S}_{k_1,k_2}\mathcal{A}_5(3,2,1,4,5)+\mathcal{S}_{k_2,k_1+k_3}\mathcal{A}_5(2,3,1,4,5) \\
\mathcal{S}_{k_3,k_4}\mathcal{A}_5(2,1,4,3,5)=&\mathcal{S}_{k_1,k_3}\mathcal{A}_5(2,3,1,4,5)+\mathcal{S}_{k_3,k_1+k_2}\mathcal{A}_5(3,2,1,4,5).
\end{align}
\end{subequations}
From the above relations we can rewrite (\ref{5KLT}) to obtain the KLT relation in the form 
\begin{align}
\mathscr{A}_5=&\frac{-1}{16\pi^2\alpha'} \bigg[ \mathcal{S}_{k_1,k_2}\mathcal{S}_{k_1,k_3}\mathcal{A}_5(1,2,3,4,5)\mathcal{A}_5(2,3,1,4,5) \nonumber \\
&+\mathcal{S}_{k_1,k_2}\mathcal{S}_{k_3,k_1+k_2}\mathcal{A}_5(1,2,3,4,5)\mathcal{A}_5(3,2,1,4,5) \bigg] \nonumber \\
&+\text{exchange of } (2 \leftrightarrow 3).
\end{align}
Similarly, we can perform the same trick but this time we consider the quadrilaterals III and IV instead. This yields the KLT relation in the form
\begin{align}
\mathscr{A}_5=&\frac{-1}{16\pi^2\alpha'} \bigg[ \mathcal{S}_{k_2,k_4}\mathcal{S}_{k_3,k_4}\mathcal{A}_5(1,2,3,4,5)\mathcal{A}_5(1,4,2,3,5) \nonumber \\
&+\mathcal{S}_{k_3,k_4}\mathcal{S}_{k_2,k_3+k_4}\mathcal{A}_5(1,2,3,4,5)\mathcal{A}_5(1,4,3,2,5) \bigg] \nonumber \\
&+\text{exchange of } (2 \leftrightarrow 3).
\end{align}
This is not a surprising result since these equivalent forms of the KLT relations were presented in the original paper \cite{KLT}. However, Plahte diagrams give us a simple geometrical way of obtaining them.

Although we deduced the geometrical expression for the KLT relations from the specific combined Plahte diagram for figure \ref{combined}, the relation (\ref{GeoKLT}) is generally valid for any of the combined pictures. In other words, we can find the KLT relations from any combined diagram not just the one presented in figure \ref{combined}, using the relation (\ref{GeoKLT}). The newly obtained KLT relations will be of the same form as equation (\ref{5KLT}) but with particles' labels interchanged. 

In deriving the KLT relations, we are free to fix the positions of any three vertices positions in the integral representation of the closed string amplitude. Then, the integral is factorised into products of open string amplitudes. In the original paper \cite{KLT}, the fixing choice, $z_1=0$, $z_4=1$ and $z_5=\infty$ was utilised for five particle scattering. 
Choosing different vertex positions to be fixed would result in exactly the KLT relations in equation (\ref{5KLT}) but with different particles' labels interchanged.

As there are fifteen different Plahte diagrams for 5-point amplitudes, it suggests that fifteen different KLT relations can be obtained from reordering the particles 1 to 5 as well. Obviously, there are 5! ways to rearrange five objects. However, not all of them corresponds to distinct KLT relations as some permutations do not change the form of relations. It is not hard to see that these following permutations keep the KLT relation in (\ref{5KLT}) invariant: First, swapping particle 1 and 4, second, swapping particle 2 and 3, and third, swapping particle 1 and 2 together with particle 3 and 4 simultaneously. As a result, the total number of different KLT relations are $5!/(2^3)=15$ as claimed.

Let us explore some more aspects of the combined diagrams in figure (\ref{combined}). One may notice that there are only eight partial amplitudes (excluding their reflections) taking part in the diagram. In order to include all twelve color-ordered amplitudes we need to enlarge the diagram. The extended version of the combined diagram in figure (\ref{combined}) is shown below in figure (\ref{extended combined}). The central quadrilaterals in figure (\ref{combined}) are disassembled into a cross-like structure in our new picture. Notice that quadrilaterals $\square$BCDE, $\square$AFCD, $\square$ABGD, and $\square$ABCH in figure (\ref{combined}) are the same quadrilaterals $\square$OBGC, $\square$OAFB, $\square$OAED, and $\square$OCHD in figure (\ref{extended combined}) respectively.

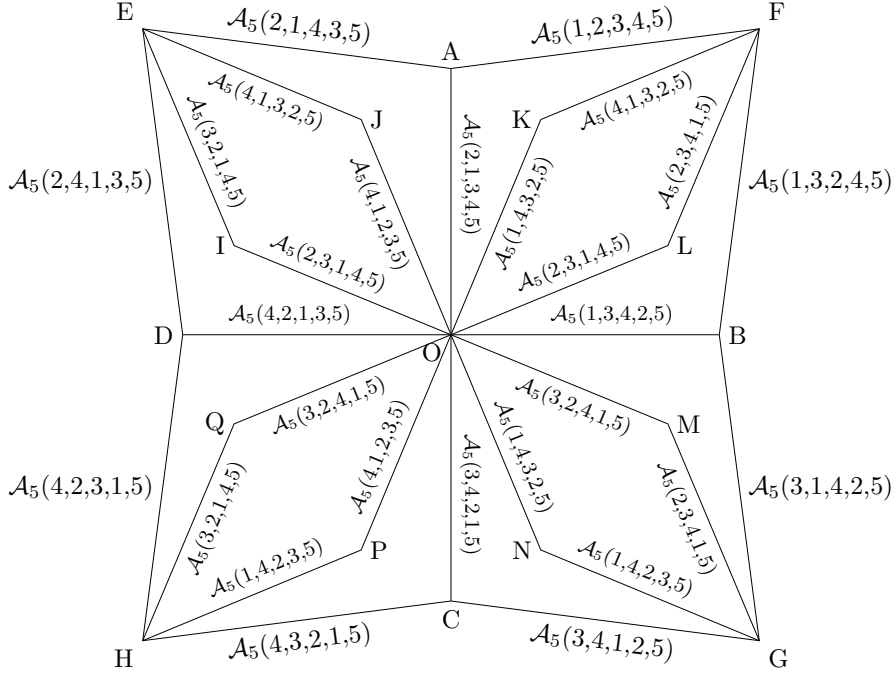
\begin{figure}[htb!]
\centering
\resizebox{0.8\linewidth}{!}{
\begin{tikzpicture}
\path (0,0) coordinate (O);
\path (90:4) coordinate (A);
\path (0:4) coordinate (B);
\path (-90:4) coordinate (C);
\path (180:4) coordinate (D);
\path (135:6.5) coordinate (E);
\path (45:6.5) coordinate (F);
\path (-45:6.5) coordinate (G);
\path (225:6.5) coordinate (H);
\draw (A)--(O) node[pos=0.4,sloped,above] {\footnotesize{$\mathcal{A}_5$(2,1,3,4,5)}}--(B) node[pos=0.6,sloped,above] {\footnotesize{$\mathcal{A}_5$(1,3,4,2,5)}} (O)--(C) node[pos=0.6,sloped,above] {\footnotesize{$\mathcal{A}_5$(3,4,2,1,5)}} (O)--(D) node[pos=0.6,sloped,above] {\footnotesize{$\mathcal{A}_5$(4,2,1,3,5)}};
\draw (B)--(F) node[pos=0.5,right] {$\mathcal{A}_5$(1,3,2,4,5)}--(A) node[pos=0.5,sloped,above] {$\mathcal{A}_5$(1,2,3,4,5)};
\draw (A)--(E)node[pos=0.5,sloped,above] {$\mathcal{A}_5$(2,1,4,3,5)} --(D) node[pos=0.5,left] {$\mathcal{A}_5$(2,4,1,3,5)};
\draw (B)--(G) node[pos=0.5,right] {$\mathcal{A}_5$(3,1,4,2,5)}--(C) node[pos=0.5,sloped,below] {$\mathcal{A}_5$(3,4,1,2,5)};
\draw (C)--(H) node[pos=0.5,sloped,below] {$\mathcal{A}_5$(4,3,2,1,5)}--(D)node[pos=0.5,left] {$\mathcal{A}_5$(4,2,3,1,5)} ;

\path (22.5:3.5) coordinate (L);
\path (67.5:3.5) coordinate (K);
\path (112.5:3.5) coordinate (J);
\path (157.5:3.5) coordinate (I);
\path (202.5:3.5) coordinate (Q);
\path (247.5:3.5) coordinate (P);
\path (292.5:3.5) coordinate (N);
\path (337.5:3.5) coordinate (M);

\draw (O)--(L)node[pos=0.6,sloped,above] {\footnotesize{$\mathcal{A}_5$(2,3,1,4,5)}}--(F)node[pos=0.4,sloped,above] {\footnotesize{$\mathcal{A}_5$(2,3,4,1,5)}};
\draw (O)--(K) node[pos=0.6,sloped,below] {\footnotesize{$\mathcal{A}_5$(1,4,3,2,5)}}--(F) node[pos=0.4,sloped,below] {\footnotesize{$\mathcal{A}_5$(4,1,3,2,5)}};
\draw (O)--(J)node[pos=0.6,sloped,below] {\footnotesize{$\mathcal{A}_5$(4,1,2,3,5)}}--(E)node[pos=0.4,sloped,below] {\footnotesize{$\mathcal{A}_5$(4,1,3,2,5)}};
\draw (O)--(I)node[pos=0.6,sloped,above] {\footnotesize{$\mathcal{A}_5$(2,3,1,4,5)}}--(E)node[pos=0.4,sloped,above] {\footnotesize{$\mathcal{A}_5$(3,2,1,4,5)}};
\draw (O)--(Q)node[pos=0.6,sloped,below] {\footnotesize{$\mathcal{A}_5$(3,2,4,1,5)}}--(H)node[pos=0.4,sloped,below] {\footnotesize{$\mathcal{A}_5$(3,2,1,4,5)}};
\draw (O)--(P)node[pos=0.6,sloped,above] {\footnotesize{$\mathcal{A}_5$(4,1,2,3,5)}}--(H)node[pos=0.4,sloped,above] {\footnotesize{$\mathcal{A}_5$(1,4,2,3,5)}};
\draw (O)--(M)node[pos=0.6,sloped,below] {\footnotesize{$\mathcal{A}_5$(3,2,4,1,5)}}--(G)node[pos=0.4,sloped,below] {\footnotesize{$\mathcal{A}_5$(2,3,4,1,5)}};
\draw (O)--(N)node[pos=0.6,sloped,above] {\footnotesize{$\mathcal{A}_5$(1,4,3,2,5)}}--(G)node[pos=0.4,sloped,above] {\footnotesize{$\mathcal{A}_5$(1,4,2,3,5)}};

\node[below left] at (O) {O};
\node[above] at (A) {A};
\node[right] at (B) {B};
\node[below] at (C) {C};
\node[left] at (D) {D};
\node[above left] at (E) {E};
\node[above right] at (F) {F};
\node[below right] at (G) {G};
\node[below left] at (H) {H};
\node[left] at (I) {I};
\node[right] at (J) {J};
\node[left] at (K) {K};
\node[right] at (L) {L};
\node[right] at (M) {M};
\node[left] at (N) {N};
\node[right] at (P) {P};
\node[left] at (Q) {Q};
\end{tikzpicture}
}
\caption{An extended version of combined Plahte diagram containing all possible color-ordered scattering amplitudes }
\label{extended combined}
\end{figure}

As we are able to combine all diagrams into one figure in the complex plane, it is natural to say that we can express all 5-point amplitudes in terms of any two amplitudes by implementing simply Euclidean geometry. Our choice is $\mathcal{A}_5(1,2,3,4,5)$ and $\mathcal{A}_5(1,3,2,4,5)$ and for convenience we call them $\mathcal{A}_1$ and $\mathcal{A}_2$ respectively. Let start by considering the quadrilateral $\square$OAFB. The vector sum of the displacements along the sides vanishes, ${\bf OA}+{\bf AF}+{\bf FB}+{\bf BO}={\bf 0}$. Taking the vector product of this relation with $\bf {OA}$ and then with ${\bf BO}$ leads to 
\begin{align*}
\mathcal{S}_{k_2,k_5} \mathcal{A}_5(1,3,4,2,5)&=\mathcal{S}_{k_1,k_2}\mathcal{A}_1+ \mathcal{S}_{k_2,k_1+k_3}\mathcal{A}_2\nonumber \\
\mathcal{S}_{k_2,k_5} \mathcal{A}_5(2,1,3,4,5)&=\mathcal{S}_{k_2,k_3+k_4}\mathcal{A}_1+\mathcal{S}_{k_2,k_4}\mathcal{A}_2.
\end{align*}

The remaining relations can be obtained by implementing a similar approach to the different quadrilaterals. This yields

\begin{align}
\mathcal{S}_{k_2,k_5}&\mathcal{S}_{k_1,k_4}\mathcal{A}(2,3,1,4,5)= (-1)^{l+1}\mathcal{S}_{k_1,k_2}\mathcal{S}_{k_3,k_4}\mathcal{A}_1 - \mathcal{S}_{k_2,k_4} \mathcal{S}_{k_1,k_3+k_4}\mathcal{A}_2 \nonumber \\
\mathcal{S}_{k_3,k_5}&\mathcal{S}_{k_1,k_4}\mathcal{A}(1,4,2,3,5)= (-1)^{l+1}\mathcal{S}_{k_1,k_2}\mathcal{S}_{k_3,k_4}\mathcal{A}_1 - \mathcal{S}_{k_1,k_3} \mathcal{S}_{k_4,k_1+k_2}\mathcal{A}_2 \nonumber \\
\mathcal{S}_{k_1,k_4}\mathcal{S}_{k_2,k_5}&\mathcal{S}_{k_3,k_5}\mathcal{A}(2,1,4,3,5)= (-1)^{l} \mathcal{S}_{k_1,k_3}\mathcal{S}_{k_2,k_4} \mathcal{S}_{k_5,k_2+k_3}\mathcal{A}_2 \nonumber \\ &+
(\mathcal{S}_{k_2,k_3+k_4}\mathcal{S}_{k_3,k_1+k_2}\mathcal{S}_{k_1,k_4}+(-1)^{l+1}\mathcal{S}_{k_2,k_3}\mathcal{S}_{k_1,k_2}\mathcal{S}_{k_3,k_4})\mathcal{A}_1
\label{minimal basis}
\end{align}
where the remaining five amplitudes are obtained by exchanging labels $2 \leftrightarrow 3$. The integer $l$ is the identifying number defined previously. This result agrees with \cite{Minimal,Mixed} which explicitly computes the expansion of color-ordered 5-point gauge amplitudes in terms of a minimal choice of two amplitudes, i.e. $\mathcal{A}_1$ and $\mathcal{A}_2$.  

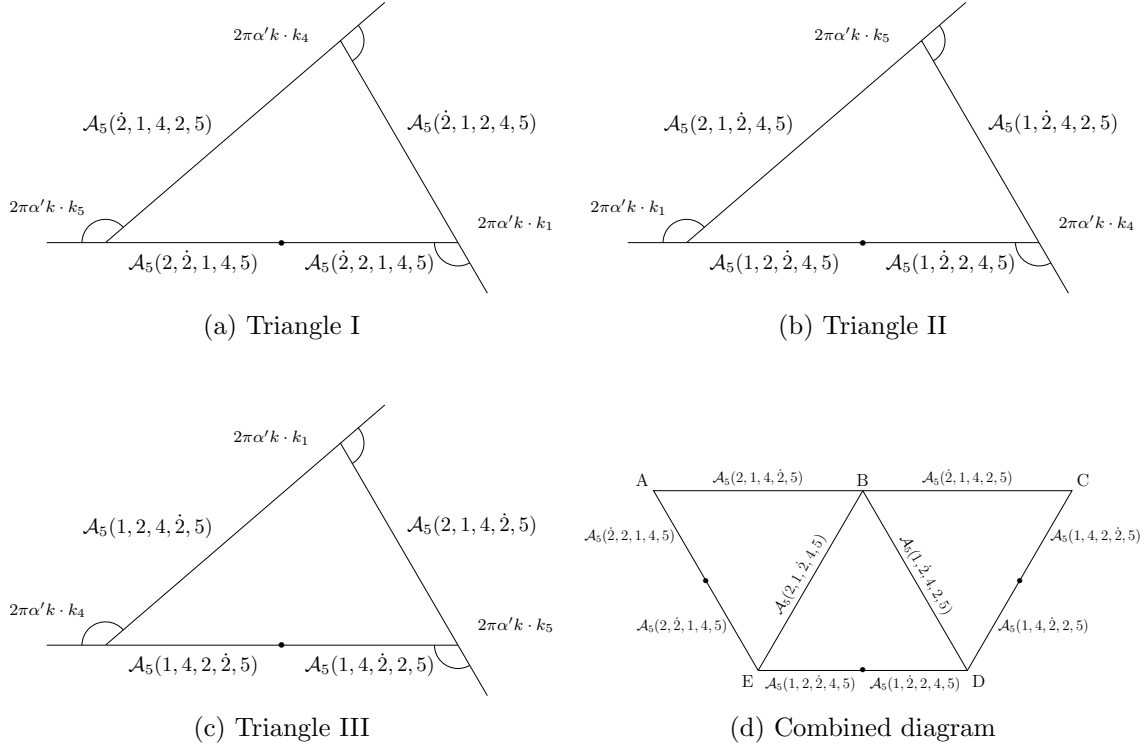
\begin{figure}[t]
\begin{subfigure}[b]{0.5\textwidth}
\centering
\resizebox{\linewidth}{!}{
\begin{tikzpicture}
\draw (0,0)--(0:6);
\path (0:6) coordinate (P);
\path (P)++(-60:1) coordinate (P1);
\path (P)++(120:4) coordinate (Q);
\path (Q)++(41:1) coordinate (Q1);
\draw (P)--(Q)--(0,0);
\draw (0,0)--(180:1) (P)--(P1) (Q)--(Q1);
\draw (0,0)++(41:.4) arc (41:180:0.4);
\draw (P)++(180:.4) arc (180:300:0.4);
\draw (Q)++(-60:.4) arc (-60:41:0.4);
\node at (3,0) {\tiny\textbullet};
\node[below] at (0:1.5) {\small{$\mathcal{A}_5(2,\dot{2},1,4,5)$}};
\node[below] at (0:4.5) {\small{$\mathcal{A}_5(\dot{2},2,1,4,5)$}};
\node[above left] at (41:2.6) {\small{$\mathcal{A}_5(\dot{2},1,4,2,5)$}};
\path (P)++(120:2) coordinate (P2);
\node[above right] at (P2) {\small{$\mathcal{A}_5(\dot{2},1,2,4,5)$}};
\node[above left] at (120:.4) {\scriptsize{$2\pi\alpha'k\cdot k_5$}};
\path (P)++(60:.4) coordinate (PP);
\node[right] at (PP) {\scriptsize{$2\pi\alpha'k\cdot k_1$}};
\path (Q)++(165:.4) coordinate (QQ);
\node[left] at (QQ) {\scriptsize{$2\pi\alpha'k\cdot k_4$}};
\end{tikzpicture}
}
\caption{Triangle I}
\vspace{.3in}
\end{subfigure}
\begin{subfigure}[b]{0.5\textwidth}
\centering
\resizebox{\linewidth}{!}{
\begin{tikzpicture}
\draw (0,0)--(0:6);
\path (0:6) coordinate (P);
\path (P)++(-60:1) coordinate (P1);
\path (P)++(120:4) coordinate (Q);
\path (Q)++(41:1) coordinate (Q1);
\draw (P)--(Q)--(0,0);
\draw (0,0)--(180:1) (P)--(P1) (Q)--(Q1);
\draw (0,0)++(41:.4) arc (41:180:0.4);
\draw (P)++(180:.4) arc (180:300:0.4);
\draw (Q)++(-60:.4) arc (-60:41:0.4);
\node at (3,0) {\tiny\textbullet};
\node[below] at (0:1.5) {\small{$\mathcal{A}_5(1,2,\dot{2},4,5)$}};
\node[below] at (0:4.5) {\small{$\mathcal{A}_5(1,\dot{2},2,4,5)$}};
\node[above left] at (41:2.6) {\small{$\mathcal{A}_5(2,1,\dot{2},4,5)$}};
\path (P)++(120:2) coordinate (P2);
\node[above right] at (P2) {\small{$\mathcal{A}_5(1,\dot{2},4,2,5)$}};
\node[above left] at (120:.4) {\scriptsize{$2\pi\alpha'k\cdot k_1$}};
\path (P)++(60:.4) coordinate (PP);
\node[right] at (PP) {\scriptsize{$2\pi\alpha'k\cdot k_4$}};
\path (Q)++(165:.4) coordinate (QQ);
\node[left] at (QQ) {\scriptsize{$2\pi\alpha'k\cdot k_5$}};
\end{tikzpicture}
}
\caption{Triangle II}
\vspace{.3in}
\end{subfigure}

\begin{subfigure}[b]{0.5\textwidth}
\centering
\resizebox{\linewidth}{!}{
\begin{tikzpicture}
\draw (0,0)--(0:6);
\path (0:6) coordinate (P);
\path (P)++(-60:1) coordinate (P1);
\path (P)++(120:4) coordinate (Q);
\path (Q)++(41:1) coordinate (Q1);
\draw (P)--(Q)--(0,0);
\draw (0,0)--(180:1) (P)--(P1) (Q)--(Q1);
\draw (0,0)++(41:.4) arc (41:180:0.4);
\draw (P)++(180:.4) arc (180:300:0.4);
\draw (Q)++(-60:.4) arc (-60:41:0.4);
\node at (3,0) {\tiny\textbullet};
\node[below] at (0:1.5) {\small{$\mathcal{A}_5(1,4,2,\dot{2},5)$}};
\node[below] at (0:4.5) {\small{$\mathcal{A}_5(1,4,\dot{2},2,5)$}};
\node[above left] at (41:2.6) {\small{$\mathcal{A}_5(1,2,4,\dot{2},5)$}};
\path (P)++(120:2) coordinate (P2);
\node[above right] at (P2) {\small{$\mathcal{A}_5(2,1,4,\dot{2},5)$}};
\node[above left] at (120:.4) {\scriptsize{$2\pi\alpha'k\cdot k_4$}};
\path (P)++(60:.4) coordinate (PP);
\node[right] at (PP) {\scriptsize{$2\pi\alpha'k\cdot k_5$}};
\path (Q)++(165:.4) coordinate (QQ);
\node[left] at (QQ) {\scriptsize{$2\pi\alpha'k\cdot k_1$}};
\end{tikzpicture}
}
\caption{Triangle III}
\end{subfigure}
\begin{subfigure}[b]{0.5\textwidth}
\centering
\resizebox{\linewidth}{!}{
\begin{tikzpicture}
\path (0:6) coordinate (P);
\path (P)++(120:6) coordinate (Q);
\path (Q)++(0:6) coordinate (R);
\path (Q)++(180:6) coordinate (S);
\draw (0,0)--(P)--(Q) node[pos=0.5, sloped, above] {\large{$\mathcal{A}_5(1,\dot{2},4,2,5)$}};
\draw (Q)--(0,0) node[pos=0.5, sloped, above] {\large{$\mathcal{A}_5(2,1,\dot{2},4,5)$}};
\draw (Q)--(R)  node[pos=0.5, sloped, above] {\large{$\mathcal{A}_5(\dot{2},1,4,2,5)$}}--(P);
\draw (Q)--(S) node[pos=0.5, sloped, above] {\large{$\mathcal{A}_5(2,1,4,\dot{2},5)$}}--(0,0);
\node at (3,0) {\textbullet};
\node at ([shift=(60:3)]P) {\textbullet};
\node at ([shift=(120:3)]0,0) {\textbullet};
\node[below] at (1.5,0) {\large{$\mathcal{A}_5(1,2,\dot{2},4,5)$}};
\node[below] at (4.5,0) {\large{$\mathcal{A}_5(1,\dot{2},2,4,5)$}};
\node[left] at ([shift=(120:1.5)]0,0]) {\large{$\mathcal{A}_5(2,\dot{2},1,4,5)$}};
\node[left] at ([shift=(120:4.5)]0,0]) {\large{$\mathcal{A}_5(\dot{2},2,1,4,5)$}};
\node[right] at ([shift=(60:1.5)]P) {\large{$\mathcal{A}_5(1,4,\dot{2},2,5)$}};
\node[right] at ([shift=(60:4.5)]P) {\large{$\mathcal{A}_5(1,4,2,\dot{2},5)$}};

\node at (0,0) [below left] {\Large{E}};
\node at (Q) [above] {\Large{B}};
\node at (P) [below right] {\Large{D}};
\node at (R) [above right] {\Large{C}};
\node at (S) [above left] {\Large{A}};
\end{tikzpicture}
}
\caption{Combined diagram}
\label{combined triangle}
\end{subfigure}
\caption{Plahte diagrams for 5-point tachyonic scattering with $k_2=k_3=k$.}
\label{special triangle}
\end{figure}

Before finishing this section, let us explore another case of interest. We will now consider a special case of 5-point scattering amplitudes where the momenta of two particles are equal, says $k_2=k_3=k$. For tachyon scattering, it causes some Plahte diagrams to become degenerate as the particles 2 and 3 are now indistinguishable. In this scenario, all component quadrilaterals in the combined diagram (\ref{combined}) can be decomposed into tree triangles which are illustrated in the figure (\ref{special triangle}). Note that we use $\dot{2}$ instead of the number 3 in the amplitudes to signify this indistinguishability.

Again, combining all three triangles gives us a combined diagrams for this special case which is presented in the figure (\ref{combined triangle}). It is not hard to find that the connection between the KLT relations and the Plahte diagrams now takes form,
\begin{align}
\mathscr{A}_5 \Big|_{k_2=k_3}=&\frac{-1}{4\pi^2\alpha'} \bigg( \frac{\langle \triangle \text{ABE} \rangle \langle \triangle \text{BDE} \rangle}{(BE)^2}\bigg)=\frac{-1}{4\pi^2\alpha'} \bigg( \frac{\langle \triangle \text{BDE} \rangle \langle \triangle \text{BCD} \rangle}{(BD)^2}\bigg). \label{S5}
\end{align}
 One can notice that the partial amplitude in the denominator is actually a side shared between two triangles. 
 
The equation (\ref{S5}) does not hold for any excited state particles in general as it involves polarization vectors. Therefore, switching the order of particles 2 and 3 no longer keeps the amplitude invariant. However, if we consider the special case where the polarization vectors of particle 2 and 3 form a rank-2 symmetric traceless tensor, $\xi_{\mu\nu}$, the KLT expression in the equation (\ref{S5}) holds. This special case of a 5-point amplitude will be useful  when discussing the mixed open and closed string amplitudes in the next section.
 

\section{Mixed Open and Closed String Amplitudes}
Tree-level scattering amplitudes for processes that involve both open and closed strings have a world-sheet with the topology of a disc. This world-sheet can be conformally mapped to the upper complex half-plane $\mathcal{H}_+=\{z \in \mathbb{C} | \text{Im}(z)\geq 0  \}$. Open string vertices are inserted along the boundary of the world-sheet while closed string vertices are inserted in the bulk. Unlike pure closed string amplitudes in which there is no interaction between left- and right- moving world-sheet fields, in mixed scattering amplitudes the interaction between these modes prevents us from expanding the amplitude into a sum of products of color-ordered open string amplitudes. Instead, the amplitude involving $N_0$ open strings and $N_C$ closed strings can be mapped into a sum of color-ordered $(N_0+2N_C)$-point open string amplitudes \cite{Mixed}.

In this paper, we will only focus on the mixed amplitudes of $(N-2)$ open strings and a single closed string. The disk amplitudes in this case can be expressed by the integral,
\begin{align}
\mathcal{M}^{(N-2,1)}&(1,4,5,\ldots,N-2;p_1,p_2)= V^{-1}_{\text{CKG}} \delta(\sum_{i\in \mathbb{N}_O, i=1}^{N-2} k_i+p_1+p_2) \nonumber \\
& \times \int_{\mathcal{I}_\sigma} \prod_{i\in \mathbb{N}_0,i=1}^{N-2}dx_i \prod_{r,s\in \mathbb{N}_0,r\neq s}^{N-2} |x_r-x_s|^{2\alpha' k_r\cdot k_s} \int_{\mathcal{H}_+} d^2z (z-\bar{z})^{2\alpha' p_1\cdot p_2} \nonumber \\  &\times \prod_{i\in \mathbb{N}_0,i=1}^{N-2} (x_i-z)^{2\alpha' p_1\cdot k_i}(x_i-\bar{z})^{2\alpha' p_2\cdot k_i} \mathcal{F}_{N-2,1}(x_i,z,\bar{z}) \label{mixed amplitude}
\end{align}
where the factor $V_{\text{CKG}}$ refers to the volume of the conformal Killing group which will be canceled out by fixing any three vertex positions. The set $\mathbb{N}_O$ is \mbox{$\{1,4,5,6,\ldots,N \}$} which contains indices used for labeling the open strings. The polarization vectors are contained in the branch-free function $\mathcal{F}_{N-2,1}$. This function depends on the types of particles we consider. This expression is for the colour-ordered partial amplitude corresponding to a group factor Tr($T_1 T_4 T_5 \ldots T_{N-2}$). This gives rise to the integration region $\mathcal{I}_\sigma=\{x\in \mathbb{R}|x_1<x_4<x_5<\ldots <x_{N-2}\}$ forcing the ordering of the open string variables. The closed string momenta $p_1$ and $p_2$ are assumed to be unrelated at first.

Following Stieberger and Taylor \cite{Disk}, the integral (\ref{mixed amplitude}) takes the form
\begin{align}
\mathcal{M}^{(N-2,1)}&(1,4,5,\ldots,N-2;p_1,p_2)=\frac{i}{2} V^{-1}_{\text{CKG}} \delta(\sum_{i=1}^{N} k_i) \int_{\mathcal{I}_\sigma}  \prod_{i\in \mathbb{N}_0,i=1}^{N-2}dx_i   \nonumber \\
&\times \prod_{r,s\in \mathbb{N}_0,r\neq s}^{N-2} |x_r-x_s|^{2\alpha' k_r\cdot k_s} \int_{-\infty}^\infty dx_2 \int_{x_2}^\infty dx_3  |x_3-x_2|^{2\alpha' k_2\cdot k_3}  \Omega(x_2,x_3)
\nonumber \\ &\times  \prod_{i\in \mathbb{N}_0,i=1}^{N-2} |x_i-x_2|^{2\alpha' k_i\cdot k_2}|x_i-\bar{z}|^{2\alpha' k_i\cdot k_3} \mathcal{F}_{N-2,1}(x_i,z,\bar{z}) \Lambda(x_i,x_2,x_3)
\end{align}
where we have redefined the closed string variables $x_2\equiv z$ and $x_3 \equiv \bar{z}$ and their corresponding momenta $k_2=p_1$ and $k_3=p_2$. The functions $\Omega(x_2,x_3)$ and $\Lambda(x_i,x_2,x_3)$ are the phase factors corresponding to the appropriate branch of the integrand. They  are defined as follows:
\begin{align}
\Omega(x_2,x_3)=& e^{2\pi i \alpha' k_2 \cdot k_3\Theta(x_3-x_2)} \nonumber \\
\Lambda(x_i,x_2,x_3) =&e^{-2\pi i \alpha' k_i \cdot k_2\Theta(x_2-x_i)}e^{2\pi i \alpha' k_i \cdot k_3\Theta(x_3-x_i)}
\end{align}
with $\Theta(x_j-x_i)$ being the Heaviside step function whose value equal to 1 for $x_j>x_i$ and 0 for otherwise. Therefore, we can write this partial mixed amplitude in terms of pure open string amplitudes as
\begin{align}
\mathcal{M}^{(N-2,1)}&(1,4,5,\ldots,N-2;p_1,p_2)= \frac{i}{2}\sum_{m\in \mathbb{N}_O,m=1}^{N-2} \sum_{n\in \mathbb{N}_O,n=m+1}^{N-1} \nonumber \\
&\times\text{exp}\Bigg\{\pi i \alpha'\Bigg( s_{23}-\sum_{i\in \mathbb{N}_O,i=1}^{m} s_{i,2}+\sum_{j\in \mathbb{N}_O,j=1}^{n} s_{j,3} \Bigg) \Bigg\} \nonumber \\
&\times\mathcal{A}_N(1,4,5,\ldots,m,2,m+1,\ldots,n,3,n+1,\ldots,N) \nonumber \\
&+\frac{i}{2}\text{exp}(\pi i \alpha' s_{23})\mathcal{A}_N(2,3,1,4,5,\ldots,N). \label{disk-open}
\end{align}
Again, $s_{ij}=2k_i \cdot k_j$.

The simplest example of a mixed disk amplitude is $\mathcal{M}^{(3,1)}$ with three open strings and one closed string. In this scenario, the closed string in the mixed disk amplitude will be replaced by a pair of collinear open strings which both carry half of the closed string momentum. If we assign both open string momenta by $k_2=k_3=k$ satisfying $k^2=-l/\alpha'$ where $l$ is the identifying number we defined earlier, the closed string momentum is now $2k$. The formula (\ref{disk-open}) yields:
\begin{align}
\mathcal{M}^{(3,1)}(1,4,5;k,k)=&\frac{i}{2}\bigg(\mathcal{A}_5(2,\dot{2},1,4,5)+\mathcal{A}_5(1,2,\dot{2},4,5)+\mathcal{A}_5(1,4,2,\dot{2},5) \nonumber \\
&+e^{2\pi i \alpha'k\cdot k_1}\mathcal{A}_5(2,1,\dot{2},4,5)+e^{2\pi i \alpha'k\cdot (k_1+k_4)}\mathcal{A}_5(2,1,4,\dot{2},5)\nonumber \\
&+e^{2\pi i \alpha'k\cdot k_4}\mathcal{A}_5(1,2,4,\dot{2},5) \bigg).
\end{align}
Due to interchangeability between particles 2 and 3, we rewrite $\dot{2}$ instead of 3 in this case. Taking the real part of the above equation into consideration, this yields
\begin{align}
\mathcal{M}^{(3,1)}(1,4,5;k,k)=&-\frac{1}{2}\bigg( \mathcal{S}_{k,k_1}\mathcal{A}_5(2,1,\dot{2},4,5)-\mathcal{S}_{k,k_5}\mathcal{A}_5(2,1,4,\dot{2},5) \nonumber \\
&+\mathcal{S}_{k,k_4}\mathcal{A}_5(1,2,4,\dot{2},5) \bigg). \label{5disk-open}
\end{align}
The open string amplitudes that result from the mixed closed/open string amplitude are actually the open string amplitudes we previously considered in the special case where $k_2=k_3$. Therefore, the equation (\ref{5disk-open}) can be further simplified using the Plahte diagrams in figure (\ref{special triangle}). These (from triangle I to III) directly give us:
\begin{subequations}
\begin{align}
&\text{I)} \quad &\mathcal{S}_{k,k_5}\mathcal{A}_5(\dot{2},1,4,2,5)= \mathcal{S}_{k,k_1}\mathcal{A}_5(\dot{2},1,2,4,5)  \\
&\text{II)} \quad &\mathcal{S}_{k,k_1}\mathcal{A}_5(2,1,\dot{2},4,5)= \mathcal{S}_{k,k_4}\mathcal{A}_5(1,\dot{2},4,2,5)  \\
&\text{III)} \quad &\mathcal{S}_{k,k_4}\mathcal{A}_5(1,2,4,\dot{2},5)= \mathcal{S}_{k,k_5}\mathcal{A}_5(2,1,4,\dot{2},5). \label{triangle conditions}
\end{align}
\end{subequations}

The above relations simplify the disk amplitude in (\ref{5disk-open}) to
\begin{align}
\mathcal{M}^{(3,1)}(1,4,5;k,k)=&-\frac{1}{2}\mathcal{S}_{k,k_1}\mathcal{A}_5(2,1,\dot{2},4,5)=-\frac{1}{2}\mathcal{S}_{k,k_5}\mathcal{A}_5(2,1,4,\dot{2},5) \nonumber \\
=&-\frac{1}{2}\mathcal{S}_{k,k_4}\mathcal{A}_5(1,2,4,\dot{2},5). \label{sim disk}
\end{align}
As a result, the disk amplitude $\mathcal{M}^{(3,1)}(1,4,5;k,k)$ can be interpreted geometrically as the  height of each Plahte diagram for 5-point tachyon scattering with any two momenta set equal in figure \ref{special triangle}.

The relations (\ref{sim disk}) also provide a description for mixed graviton and gauge boson ampltitudes. The gravition in the mixed disk amplitude can be split into pairs of collinear gauge vectors. In the field theory limit, the left-hand side term in (\ref{sim disk}) is described by Einstein-Yang-Mills theory which express the decay of a graviton into three gauge bosons \cite{Bern:1999bx}.

The collinear limit for Yang-Mills amplitudes may seem troublesome for our mixed disk amplitudes. It is known that the partial amplitudes with adjacent gauge bosons contain collinear divergence \cite{Mangano:1990by}. Fortunately,
these singularities are absent from the partial amplitudes in the expression (\ref{sim disk}) as the collinear pair are not adjacent.

Furthermore, one can also make a connection between closed string amplitudes and mixed disk amplitudes. According to the equation (\ref{S5}) and (\ref{sim disk}), it is not hard to obtain
\begin{equation}
\mathscr{A}_5 \bigg|_{k_2=k_3=k}=-\bigg(\frac{1}{2\pi\alpha'}\bigg)^2 \big( \mathcal{M}^{(3,1)}(1,4,5;k,k) \big)^2.
\end{equation}
This expresses the 5-point closed tachyon string amplitudes with any two momenta being equal as a quadratic in the disk amplitudes describing the scattering of  three open string and one closed string tachyon.

More interestingly, this allows us to compute the specific case of the five-point graviton scattering amplitude as a product of the scattering amplitudes of three massless gauge bosons and a graviton. The relation is presented in tensor form as
\begin{align}
\mathscr{A}_5&^{\mu_1\nu_1\ldots\mu_5\nu_5} \bigg|_{k_2=k_3=k} = \nonumber \\&-\bigg(\frac{1}{2\pi\alpha'}\bigg)^2 \mathcal{M}^{(3,1)\mu_1\mu_2\ldots\mu_5}(1,4,5;k,k)\mathcal{M}^{(3,1)\nu_1\nu_2\ldots\nu_5}(1,4,5;k,k).
\end{align}
The symmetric traceless polarization vectors $\xi_{\mu\nu}$ are to be contracted with both sides to obtain the scattering amplitude.


\section{Comments on the Connection between Plahte Diagrams and BCFW \mbox{Recursion} Relations}

In the first decade of this century the study of scattering amplitudes benefitted considerably from the discovery of the Britto-Cachazo-Feng-Witten (BCFW) on-shell recursion relations \cite{Britto_2005,Britto2005NewRR}. The relations allow us to express  tree-level amplitudes as products of other tree-level amplitudes with fewer particles. The key idea for deriving the on-shell recursion relations is based on the fact that any tree-level scattering amplitude is a rational function of the external momenta, thus, one can turn an amplitude $A_n$ into a complex meromorphic function $A_n(z)$ by deforming the external momenta through introducing a complex variable $z$. The deformed momenta are required to be on-shell and satisfy momentum conservation. For a scattering process involving $n$ particles, we can choose an arbitrary pair of particle momenta to be shifted. Our choice is given by
\begin{subequations}
\begin{align}
    k_1 \rightarrow \hat{k}_1(z) =& k_1-qz \\
    k_n \rightarrow \hat{k}_n(z) =& k_n+qz
\end{align}
\label{shifted momenta}
\end{subequations}
where $q$ is a reference momentum which obeys $q\cdot q=k_1\cdot q=k_n\cdot q=0$. 

The unshifted amplitude $A_n(Z=0)$ can be obtained from a contour integration in which the contour is large enough to enclose all finite poles. According to Cauchy's theorem,
\begin{equation}
    A_n(0)=\oint dz \frac{A_n(z)}{z}-\sum_\text{poles} \text{Res}_{z=z_\text{poles}}. \label{cauchy}
\end{equation}
If the amplitude is well-behaved at large $z$ (which is the case for  most theories), then the amplitude at $z=0$ is equal to the sum of the residues over the finite poles. For Yang-Mills theory, the residue at a finite pole is the product of amplitudes with at least two fewer particles and one leg for an exchanged particle. In Yang-Mills a sum over the helicities of the intermediate gauge boson and in general theories a sum over all allowed intermediate particle states must also be done. In the general case case, the BCFW recursion relation is 
\begin{equation}
    A_n(0)=\sum_{\substack{\text{poles}\\ \alpha}}\sum_{\substack{\text{physical}\\ \text{states}}} A_L(\dots,P(z_\alpha))\frac{2}{P^2+M^2}A_R(-P(z_\alpha),\dots)
    \label{BCFW}
\end{equation}
with $P$ being the momentum of the exchanged particle with mass $M$. 

The validity of equation (\ref{cauchy}) requires the absence of a pole at infinity. In the case that there exists such a pole, one must include the residue at infinity. However, the residue at this pole does not have a similar physical interpretation to the residues at finite poles. A detailed discussion can be found in \cite{Feng_2010}. 

The idea of deforming scattering amplitudes can be applied to string theory as well. Despite the infinite number of physical states of intermediate particles, many works have successfully addressed the string theory versions of BCFW on-shell recursion relations  \cite{Boels:2008fc,Boels:2010bv,chang2012,Cheung_2010}. 

There are links between  Plahte diagrams and the BCFW on-shell recursion relations. We have noticed that when we collapse any two adjacent sides of a 5-point gluonic Plahte diagram to the diagonal line, the diagonal line along with two remaining partial amplitudes forms a triangle. It turns out that the corresponding Plahte identities for the triangle coincide with the BCJ relations derived from the BCFW recursion relations of the five gluon scattering amplitudes.

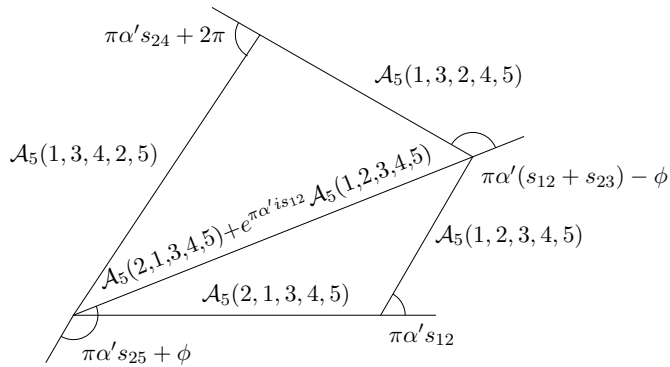
\begin{figure}[t]
\centering
\resizebox{.6\linewidth}{!}{
\begin{tikzpicture}
\path (0:5) coordinate (P);
\path (P)++(60:3) coordinate (Q);
\path (Q)++(150:4) coordinate (R);
\draw (0,0)--(P)--(Q)--(R)--cycle;
\path (-120:.9) coordinate (O1);
\path (P)++(0:.9) coordinate (P1);
\path (Q)++(21.787:.9) coordinate (Q1);
\path (R)++(150:.9) coordinate (R1);
\draw (0,0)--(O1) (P)--(P1) (Q)--(Q1) (R)--(R1);
\draw (0,0)--(Q) node[pos=0.5,sloped,above] {$\mathcal{A}_5$(2,1,3,4,5)+$e^{\pi\alpha'is_{12}}\mathcal{A}_5$(1,2,3,4,5)};
\draw ([shift=(-120:0.4)]0,0) arc(-120:21.787:0.4);
\draw ([shift=(0:0.4)]P) arc(0:60:0.4);
\draw ([shift=(21.787:0.4)]Q) arc(21.787:150:0.4);
\draw ([shift=(150:0.4)]R) arc(150:240:0.4);
\node[above] at (3.3,0) {$\footnotesize{\mathcal{A}_5(2,1,3,4,5)}$};
\node[right] at ([shift=(60:1.5)]P) {$\footnotesize{\mathcal{A}_5(1,2,3,4,5)}$};
\node[above right] at ([shift=(150:2)]Q) {$\footnotesize{\mathcal{A}_5(1,3,2,4,5)}$};
\node[above left] at ([shift=(56.5:2.76)]0,0) {$\footnotesize{\mathcal{A}_5(1,3,4,2,5)}$};
\node[below right] at (0,-0.3)
{$\scriptsize{\pi\alpha's_{25}+\phi}$};
\node[below right] at (P) {$\scriptsize{\pi\alpha's_{12}}$};
\node[ below right] at (Q) {$\scriptsize{\pi\alpha'(s_{12}+s_{23})-\phi}$};
\node[left] at ([xshift=-12]R) {$\scriptsize{\scriptsize{\pi\alpha's_{24}+2\pi}}$};
\end{tikzpicture}
}
\caption{An triangle made from the diagonal line of the five-point gluonic Plahte diagram.}
\label{diagonal triangle}
\end{figure}

As an explicit example, consider the triangle in the figure \ref{diagonal triangle}. The external angles next to the diagonal line of a triangle are parametrized by $\phi$. The parameter $\phi$ can be evaluated using BCFW on-shell recursion relations in which we will see later that it corresponds to the shifted momenta in (\ref{shifted momenta}). Without much effort, one can find the Plahte identities for the triangle as
\begin{align}
    \frac{|\mathcal{A}_5(1,3,2,4,5)|}{\sin(\pi\alpha's_{25}+\phi)}                   &=       \frac{|\mathcal{A}_5(2,1,3,4,5)+e^{\pi\alpha'is_{12}}\mathcal{A
    }_5(1,2,3,4,5)|}{\sin(\pi\alpha's_{24})}                          \nonumber \\
    &=\frac{|\mathcal{A}_5(1,3,4,2,5)|}{\sin(\pi\alpha'(s_{12}+s_{23})-\phi)}. \label{sine law diagonal}
\end{align}

For convenience, let us make a specific choice of polarizations, say negative helicity for particles one and five and positive helicity for those remaining. Let us calculate the on-shell recursion relations for $A_5(1^-,2^+,3^+,4^+,5^-)$ based on the $[1,5\rangle$-shift (the shifted momentun $q=|5\rangle[1|$). According to the equation (\ref{BCFW}), the amplitude breaks down into two terms as
\begin{align}
    A_5(1^-,2^+,3^+,4^+,5^-)=\hat{A}_3(\hat{1}^-,2^+,-\hat{P}^+_{12})\frac{1}{\hat{s}_{12}}\hat{A}_4(\hat{P}^-_{12},3^+,4^+,\hat{5}^-) \nonumber \\
    +\hat{A}_4(\hat{1}^-,2^+,3^+,\hat{P}^-_{45})\frac{1}{\hat{s}_{45}}\hat{A}_3(-\hat{P}^+_{45},4^+,\hat{5}^-). \label{BCFW1}
\end{align}
The notation $P_{ij}$ means $k_i+k_j$. Note that all hatted terms are evaluated at the residue value such that $\hat{s}_{ij}=0$ or $z=z_{ij}=-P_{ij}^2/2q\cdot P_{ij}$. Now let us take a closer look at the subamplitude $\hat{A}_3(-\hat{P}^+_{45},4^+,\hat{5}^-)$. According to the Parke-Taylor formula,
\begin{equation}
    \hat{A}_3(-\hat{P}^+_{45},4^+,\hat{5}^-)=\frac{[\hat{P}4]^3}{[4\hat{5}][\hat{5}\hat{P}]}.
\end{equation}
It turns out that all spinor products in above expression are zeroes. More detailed analysis can be found in \cite{elvang_huang_2015}. As there are three powers in the numerator compared with the two in the denominator, the subamplitude $\hat{A}_3(-\hat{P}^+_{45},4^+,\hat{5}^-)$ vanishes. Consequently, only the first term from (\ref{BCFW1}) contributes.

Using a similar approach, we can then find the remaining partial amplitudes as
\begin{align}
    A_5(1^-,3^+,2^+,4^+,5^-)=&\hat{A}_3(\hat{1}^-,3^+,-\hat{P}^+_{13})\frac{1}{\hat{s}_{13}}\hat{A}_4(\hat{P}^-_{13},2^+,4^+,\hat{5}^-) \nonumber \\
    A_5(1^-,3^+,4^+,2^+,5^-)=&\hat{A}_3(\hat{1}^-,3^+,-\hat{P}^+_{13})\frac{1}{\hat{s}_{13}}\hat{A}_4(\hat{P}^-_{13},4^+,2^+,\hat{5}^-) \nonumber \\
    A_5(2^+,1^-,3^+,4^+,5^-)=&-A_5(1^-,2^+,3^+,4^+,5^-) \nonumber \\
    &+\hat{A}_3(\hat{1}^-,3^+,-\hat{P}^+_{13})\frac{1}{\hat{s}_{13}}\hat{A}_4(\hat{P}^-_{13},4^+,\hat{5}^-,2^+).
\end{align}
Notice that these colour-ordered amplitudes  can now be related to each other if we exploit the BCJ relations for the subamplitude $\hat{A}_4(\hat{P}^-_{13},2^+,4^+,\hat{5}^-)$:

\begin{align}
    \frac{|A_5(1,3,2,4,5)|}{s_{25}+2z_{13}q\cdot k_2}                   &=       \frac{|A_5(2,1,3,4,5)+A_5(1,2,3,4,5)|}{s_{24}}                          \nonumber \\
    &=\frac{|A_5(1,3,4,2,5)|}{(s_{12}+s_{23})-2z_{13}q\cdot k_2}\label{BCJ BCFW}
\end{align}
where $z_{13}=-P_{13}^2/2q\cdot P_{13}$. 

Clearly, The BCJ relations (\ref{BCJ BCFW}) resemble the field theory version of the Plahte identities in the equation (\ref{sine law diagonal}) with $\phi=2z_{13}q\cdot k_2$. The parameter $\phi$ which refers to the amount angles are shifted by is now related to the shifted momentum $zq$ from BCFW recursion relations as claimed. For other sets of polarizations, the shifted angles can be obtained in a similar manner but with different choices of shifted momenta. 


\section{Plahte Diagrams with Complex Momenta}
String scattering amplitudes considered as mathematical objects have been widely studied in past few decades. For example, they provide a close connection to local zeta functions especially in the framework of $p$-adic string theory \cite{FREUND1987191,2016arXiv161103807B,Garcia-Compean:2019jvk,Bocardo-Gaspar:2017atv}. Recently, the work of Bocardo-Gaspar, Veys and Z{\'u}{\~n}iga-Galindo \cite{2019arXiv190510879B}  established in a rigorous mathematical way that the integral expressions for open string amplitudes (\ref{open amp}) are bona fide integrals which admit meromorphic continuations as complex functions in the kinematic parameters.

When the momenta $k_i$ are taken to be complex the Plahte diagram is deformed. External angles between the sides representing amplitudes are shifted by the differences of the internal phases of the corresponding amplitudes. Besides, the partial amplitudes themselves are re-scaled due to the presence of the imaginary component of the kinematic variables $k_i\cdot k_j$ .

Let us consider the generalization of the Plahte identity for $n$ particles scattering. As the momenta are allowed to become complex, the amplitudes also become complex so we write them in Euler's form as $\mathcal{A}_n(\sigma)=|\mathcal{A}_n(\sigma)|e^{i\varphi_\sigma}$ where $\sigma$ is a certain ordering of open string vertices. In this scenario, the Plahte identity in equation (\ref{Plahte}) becomes
\begin{align}
|\mathcal{A}_n&(2,1,3,\ldots,n)|e^{i\varphi_1}+|\mathcal{A}_n(1,2,3,\ldots,n)|e^{-\pi i\alpha's_{12}+i\varphi_3}\nonumber \\& +|\mathcal{A}_n(1,3,2,\ldots,n)|e^{-\pi i\alpha'(s_{12}+s_{23})+i\varphi_4}
\nonumber \\& +\ldots+|\mathcal{A}_n(1,3,\ldots,n-1,2,n)|e^{-\pi i\alpha'(s_{12}+s_{23}+\ldots+s_{2(n-1)})+i\varphi_n}=0 \label{complex Plahte}
\end{align}
where $s_{ij}=2k_i\cdot k_j$. The internal phase $\varphi_i$
is labelled by the particle ordering with particle $i$ next to the particle 2 to its right. Note that the complex momenta $k_i$ are constrained by  $\sum_{i=1}^n k_i=0$ and $k_i\cdot k_i=-l/\alpha'$ where $l=-1$ and $0$ for tachyons and gauge bosons respectively.

\begin{figure}[t]
\centering
\begin{subfigure}{.9\textwidth}
\resizebox{\linewidth}{!}{
\begin{tikzpicture}
\path (0,0) coordinate (A);
\path (0:4) coordinate (B);
\path (B)++(72:4) coordinate (C);
\path (C)++(144:4) coordinate (D);
\path (D)++(216:4) coordinate (E);
\path (B)++(0:1) coordinate (B1);
\path (C)++(72:1) coordinate (C1);
\path (D)++(144:1) coordinate (D1);
\path (E)++(216:1) coordinate (E1);
\path (A)++(288:1) coordinate (A1);

\draw (A)--(B) node[pos=0.5,above] 
{$|\mathcal{A}_n(2,1,3,\ldots,n)|$}--(C) node[pos=0.5,right] {$|\mathcal{A}_n(1,2,3,\ldots,n)|D(\text{Im}(s_{12}))$}--(D) node[pos=0.6, above right] {$|\mathcal{A}_n(1,3,2,\ldots,n)|D(\text{Im}(s_{12}+s_{23}))$} (E)--(A)node[pos=0.5,left] {$|\mathcal{A}_n(1,3,4,\ldots,2,n)|D^{-1}(\text{Im}(s_{2n}))$};
\draw (B)--(B1) (C)--(C1) (D)--(D1) (E)--(E1) (A)--(A1);
\path ([shift=(36:1)]E) coordinate (E2);
\path ([shift=(216:1)]D) coordinate (D2);
\draw (E)--(E2);
\draw (D)--(D2);
\path (E2) -- (D2) node[midway] [sloped] {\ldots};
\draw (A)++(-72:0.5) arc (-72:0:0.5);
\draw (B)++(0:0.5) arc (0:72:0.5);
\draw (C)++(72:0.5) arc (72:144:0.5);
\draw (E)++(216:0.5) arc (216:288:0.5);
\node at ([shift=(-32:.5)]A) [below right] {$-\pi \alpha'\text{Re}( s_{2n})+\Delta\varphi_{1n}$};
\node at ([shift=(36:.5)]B) [above right] {$-\pi \alpha'\text{Re}( s_{12})+\Delta\varphi_{31}$};
\node at (C) [right] {$-\pi \alpha' \text{Re}(s_{23})+\Delta\varphi_{43}$};
\node at (E) [above left] {$-\pi \alpha'\text{Re}( s_{2(n-1)})+\Delta\varphi_{n(n-1)}$};
\end{tikzpicture}
}
\end{subfigure}
\caption{Plahte diagram for $N$-point open tachyon string amplitudes with complex momenta corresponding to the Plahte identity (\ref{complex Plahte})}
\label{complex N Plahte}
\end{figure}
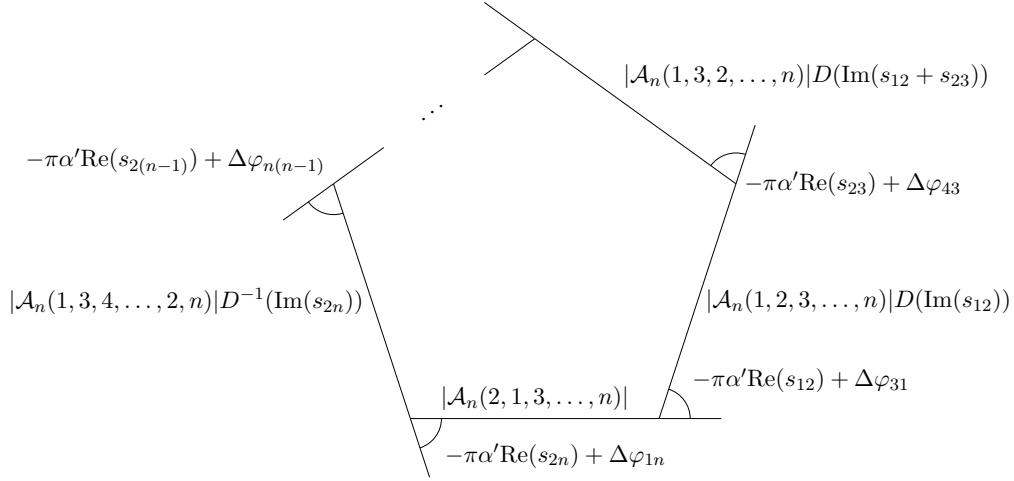

According to the Plahte identity (\ref{complex Plahte}), it is not hard to see that the internal phases $\varphi_i$ alter the external angles in the Plahte diagram and also that the imaginary components of kinematic variables Im($s_{ij}$) lead to a re-scaling of the sides of the diagram.

The Plahte diagram corresponding to the Plahte identity (\ref{complex Plahte}) is presented in figure \ref{complex N Plahte}. The function $D(x)$ is a scaling factor defined as $e^{\pi\alpha'x}$ where $x$ is real. $\Delta\varphi_{ij}\equiv \varphi_i-\varphi_j$ is an internal phase difference. The Plahte diagram presented is a generalisation of that of  figure \ref{N Plahte}.

Recall that in order to get the identity (\ref{complex Plahte}), the integral (\ref{Start integral}) was performed along the contour which is closed in the upper-half plane. Another identity can be found using the same integral but with the contour closed in the lower-half plane instead. This yields 
\begin{align}
|\mathcal{A}_n&(2,1,3,\ldots,n)|e^{i\varphi_1}+|\mathcal{A}_n(1,2,3,\ldots,n)|e^{\pi i\alpha's_{12}+i\varphi_3}\nonumber \\& +|\mathcal{A}_n(1,3,2,\ldots,n)|e^{\pi i\alpha'(s_{12}+s_{23})+i\varphi_4}
\nonumber \\& +\ldots+|\mathcal{A}_n(1,3,\ldots,n-1,2,n)|e^{\pi i\alpha'(s_{12}+s_{23}+\ldots+s_{2(n-1)})+i\varphi_n}=0. \label{complex conjugate Plahte}
\end{align}

\begin{figure}[thb!]
\centering
\begin{subfigure}{.9\textwidth}
\resizebox{\linewidth}{!}{
\begin{tikzpicture}
\path (0,0) coordinate (A);
\path (0:4) coordinate (B);
\path (B)++(72:4) coordinate (C);
\path (C)++(144:4) coordinate (D);
\path (D)++(216:4) coordinate (E);
\path (B)++(0:1) coordinate (B1);
\path (C)++(72:1) coordinate (C1);
\path (D)++(144:1) coordinate (D1);
\path (E)++(216:1) coordinate (E1);
\path (A)++(288:1) coordinate (A1);

\draw (A)--(B) node[pos=0.5,above] 
{$|\mathcal{A}_n(2,1,3,\ldots,n)|$}--(C) node[pos=0.5,right] {$|\mathcal{A}_n(1,2,3,\ldots,n)|D^{-1}(\text{Im}(s_{12}))$}--(D) node[pos=0.6, above right] {$|\mathcal{A}_n(1,3,2,\ldots,n)|D^{-1}(\text{Im}(s_{12}+s_{23}))$} (E)--(A)node[pos=0.5,left] {$|\mathcal{A}_n(1,3,4,\ldots,2,n)|D(\text{Im}(s_{2n}))$};
\draw (B)--(B1) (C)--(C1) (D)--(D1) (E)--(E1) (A)--(A1);
\path ([shift=(36:1)]E) coordinate (E2);
\path ([shift=(216:1)]D) coordinate (D2);
\draw (E)--(E2);
\draw (D)--(D2);
\path (E2) -- (D2) node[midway] [sloped] {\ldots};
\draw (A)++(-72:0.5) arc (-72:0:0.5);
\draw (B)++(0:0.5) arc (0:72:0.5);
\draw (C)++(72:0.5) arc (72:144:0.5);
\draw (E)++(216:0.5) arc (216:288:0.5);
\node at ([shift=(-32:.5)]A) [below right] {$-\pi \alpha'\text{Re}( s_{2n})-\Delta\varphi_{1n}$};
\node at ([shift=(36:.5)]B) [above right] {$-\pi \alpha'\text{Re}( s_{12})-\Delta\varphi_{31}$};
\node at (C) [right] {$-\pi \alpha' \text{Re}(s_{23})-\Delta\varphi_{43}$};
\node at (E) [above left] {$-\pi \alpha'\text{Re}( s_{2(n-1)})-\Delta\varphi_{n(n-1)}$};
\end{tikzpicture}
}
\end{subfigure}
\caption{Plahte diagram for $N$-point open tachyon string amplitudes with complex momenta corresponding to the Plahte identity (\ref{complex conjugate Plahte})}
\label{complex conjugate N Plahte}
\end{figure}
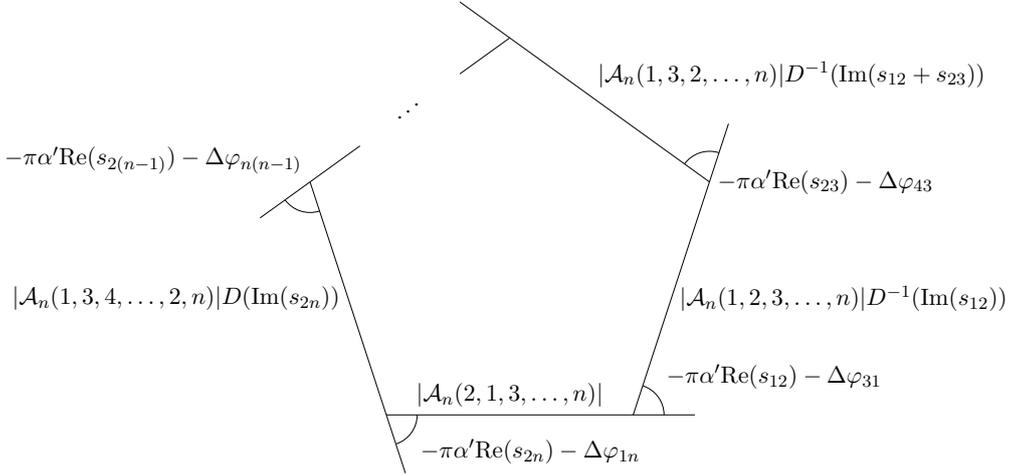

Unlike the identities with real kinematic variables, closing the contour in the  upper-half or lower-half plane generates distinct Plahte identities. It can be seen from the relations (\ref{complex Plahte}) and (\ref{complex conjugate Plahte}) that both identities provide different information. As a result, they create different Plahte diagrams. The diagram corresponding to the identity (\ref{complex conjugate Plahte}) is illustrated in figure \ref{complex conjugate N Plahte}. 

It is clear from the figures that the external angles and sides are shifted and re-scaled differently in both diagrams. However, when all imaginary parts of kinematic variables $s_{ij}$ are tuned off, both diagrams become identical.

We now give an explicit example. For simplicity, we phrase the discussion only for four-point scattering. By combining equations (\ref{complex Plahte}) and (\ref{complex conjugate Plahte}), we can find relations among the complex amplitudes as 
\begin{equation}
    \frac{|\mathcal{A}_4(1,2,3,4)|e^{i\varphi_3}}{\sin(\pi \alpha' s_{24})}=\frac{|\mathcal{A}_4(2,1,3,4)|e^{i\varphi_1}}{\sin(\pi \alpha's_{23})}=\frac{|\mathcal{A}_4(1,3,2,4)|e^{i\varphi_4}}{\sin( \pi \alpha's_{12})} \label{complex sine law}
\end{equation}
which is a complex continuation of equation (\ref{sine law}). It relates all partial amplitudes to one amplitude. The relation above also allows us to find connections between internal phases $\varphi_i$ as linear relations. By dividing the equation (\ref{complex sine law}) with its own conjugation, one can obtain
\begin{align}
&&    \varphi_1=\varphi_3-\frac{i}{2}\ln\Bigg( \frac{\sin(\pi\alpha'\bar{s}_{24})\sin(\pi\alpha's_{23})}{\sin(\pi\alpha's_{24})\sin(\pi\alpha'\bar{s}_{23})} \Bigg), \nonumber \\ \text{and} &&
        \varphi_4=\varphi_3-\frac{i}{2}\ln\Bigg( \frac{\sin(\pi\alpha'\bar{s}_{24})\sin(\pi\alpha's_{12})}{\sin(\pi\alpha's_{24})\sin(\pi\alpha'\bar{s}_{12})} \Bigg). \label{internal phases}
\end{align}
Equivalently, The relations (\ref{internal phases}) can also be expressed as 
\begin{align}
&&    \varphi_1=\varphi_3-\frac{1}{2}\arctan(\mathcal{K}(\pi\alpha's_{24}))+\frac{1}{2}\arctan(\mathcal{K}(\pi\alpha's_{23})), \nonumber \\ \text{and} &&
        \varphi_4=\varphi_3-\frac{1}{2}\arctan(\mathcal{K}(\pi\alpha's_{24}))+\frac{1}{2}\arctan(\mathcal{K}(\pi\alpha's_{12})), \label{internal phases 2}
\end{align}
where $\mathcal{K}(z)$ is defined as
\begin{equation}
    \frac{2\sin(\text{Re}(z))\cos(\text{Re}(z))\sinh(\text{Im}(z))\cosh(\text{Im}(z))}{\sin^2(\text{Re}(z))\cosh^2(\text{Im}(z))-\cos^2(\text{Re}(z))\sinh^2(\text{Im}(z))}.
\end{equation}
The above relations are valid for all types of particles.


\section{Conclusions}

To conclude, we have investigated geometrical diagrams based on linear monodromy relations between open string amplitudes  discovered by Plahte. Colour-ordered open string amplitudes and kinematic variables are represented in these diagrams as polygonal sides and external angles respectively. We have generalised the diagrams to complex momenta when the amplitudes have a  meromorphic continuation. For complex momenta, the diagrams are deformed such that external angles are shifted by the difference between internal phases of adjacent amplitudes and the sides themselves are re-scaled based on the imaginary components of kinematic variables.
 
The Plahte diagrams for five-particle scattering are depicted as quadrilaterals. By combining different quadrilaterals together, we were able to  express the KLT relations  relating closed and open string amplitudes for five-point scattering as geometrical expressions. Furthermore, we used the diagrams to re-derive the fact that all five-point amplitudes can be expressed in terms of two selected amplitudes \cite{Minimal,Mixed}.

Mixed open and closed string amplitudes were also investigated using the geometrical expression of the KLT relations. It was found that the five-point closed tachyon string amplitudes with any two momenta set equal can be expressed as a quadratic in the disk amplitudes describing three open string and one closed string. This result holds for all excited states.

Finally, we described a  connection between Plahte diagrams and BCFW on-shell recursion relations. We  noticed that a triangle obtained from a diagonal line of the diagrams for five-gluon scattering coincides with the BCJ relations derived from the BCFW recursion relations of the five-gluon scattering amplitudes.

\acknowledgments We are pleased to acknowledge Development and Promotion of Science and Technology Talents Project (Royal Thai Government Scholarship) for support.

\bibliographystyle{unsrt}
\bibliography{ref.bib}

\newpage

\begin{appendices}
\section{Plahte Identities} \label{appen Plahte identities}

We will here briefly review the derivation of the Plahte identities. The general expression for an $n$-point color-ordered open string amplitude is \cite{Green}

\begin{align}
\mathcal{A}_n(1,2,\ldots,n)=&\int \prod_{i=1}^{n} dz_i \frac{|z_{ab}z_{bc}z_{ac}|}{dz_a dz_b dz_c} \prod_{i=1}^{n-1} \Theta(x_{i+1}-x_i) \nonumber \\  &\times \prod_{1\leq i < j \leq n} |z_{ij}|^{2\alpha' k_i \cdot k_j} \mathcal{F}_n \label{open amp}
\end{align}
where $\Theta(x_j-x_i)$ is the Heaviside step function forcing the ordering of external legs as $x_j>x_i$ since $\Theta(x)=1$ for $x\geq 0$ and $\Theta(x)=0$ for otherwise. For the bosonic string, $dz_i=dx_i$ and $z_{ij}=x_i-x_j$ while for the superstring case $dz_i=dx_id\theta_i$ and $z_{ij}=x_i-x_j+\theta_i\theta_j$. Invariance under M\"{o}bius transformations allows us to set any three arbitrary integration variables, i.e. $z_a$, $z_b$ and $z_c$ equal to any fixed distinct values. A conventional choice is $x_1=0$, $x_{n-1}=1$ and $x_n=+\infty$ for the bosonic string as well as $\theta_{n-1}=\theta_{n}=0$ for the supersymmetric case. 

The function $\mathcal{F}_n$ is a branch free function which comes from the operator product expansion of vertex operators depending on the external states of the amplitude we consider. $\mathcal{F}_n=1$ for tachyons and $\mathcal{F}_n=\text{exp}\Big( \sum_{i>j}\frac{\xi_i\cdot \xi_j}{(x_i-x_j)^2}-\sum_{i\neq j}\frac{k_i\cdot \xi_j}{(x_i-x_j)}  \Big)\Big|_\text{multilinear in $\xi_i$}$ for an $n$-gauge field amplitude with $n$ polarization vectors $\xi_i$. In addition, $\mathcal{F}_n=\int\prod_{i=1}^{n}d\eta_i \allowbreak \times \text{exp}\Big[\sum_{i\neq j} \Big(  \frac{\sqrt{\alpha'}\eta_i(\theta_i-\theta_j)(\xi_i\cdot k_j)-\eta_i\eta_j(\xi_i\cot \xi_j)}{(x_i-x_j+\theta_i\theta_j)}\Big) \Big]$ for the superstring amplitude where $\eta_i$ are Grassmann variables.

Consider the complex integral
\begin{align}
\int^{\infty}_{-\infty} dx_2 \int \prod_{i=3}^{n-2} dx_i  &\bigg( \Theta(x_3-x_1) \prod_{i=3}^{n-1} \Theta(x_{i+1}-x_i) \nonumber \\  &\times \prod_{1\leq i < j \leq n} (x_j-x_i)^{2\alpha' k_i \cdot k_j} \mathcal{F}_n \bigg) \label{Start integral}
\end{align}
where we choose $x_1=0$, $x_{n-1}=1$ and $x_n=+\infty$. The ordering of variables $x_i$ is $x_1<x_3<x_4<\ldots<x_{n-1}<x_n$ as a result of the Heaviside step functions. The integrand in (\ref{Start integral}) contains $n-2$ branch points with respect to $x_2$ where all branch points are situated along the real axis. The integration with respect to the variable $x_2$ can be performed slightly above the real axis to avoid the branch points and can then be closed in the upper half plane. The integral vanishes due to the absence of singularities. To relate the term $(x_j-x_i)^{2\alpha' k_i \cdot k_j}$ 
to the $|z_{ij}|^{2\alpha' k_i \cdot k_j}$ in (\ref{open amp}) the following relations are useful
\begin{equation}
x^c=
\begin{cases}
e^{i\pi c}(-x)^c, & \text{Im}(x)\geq 0 \\
e^{-i\pi c}(-x)^c, & \text{Im}(x)< 0
\end{cases} \label{z-z}
\end{equation}
when $x<0$. As a result, we can obtain the Plahte identity for bosonic string as 
\begin{align}
e^{2\pi i\alpha'k_1\cdot k_2}&\mathcal{A}_n(2,1,3,\ldots,n)+\mathcal{A}_n(1,2,3,\ldots,n)+e^{-2\pi i\alpha'k_2\cdot k_3}\mathcal{A}_n(1,3,2,\ldots,n) \nonumber \\&
+\ldots+e^{-2\pi i\alpha'k_2\cdot (k_3+k_4+\ldots+k_{n-1})}\mathcal{A}_n(1,3,\ldots,n-1,2,n)=0. \label{Plahte}
\end{align}
The other Plahte identities can be found by a similar approach but using different orderings and integration variables. Note that all the amplitudes appearing in a Plahte identity involve the same states and polarizations.

The complex conjugate relation can be obtained using a similar contour in the lower half-plane. The combination of these  two identities leads to the following relations:
\begin{align}
0=&\mathcal{A}_n(2,1,3,\ldots,n)+\mathcal{C}_{k_1,k_2}\mathcal{A}_n(1,2,3,\ldots,n)\nonumber \\
&+\mathcal{C}_{k_2,k_1+k_3}\mathcal{A}_n(1,3,2,\ldots,n) \nonumber \\
&+\ldots+\mathcal{C}_{k_2,k_1+k_3+\ldots+k_{n-1}}\mathcal{A}_n(1,3,4,\ldots,n-1,2,n) \label{real Plahte}
\end{align}
and
\begin{align}
0=&\mathcal{S}_{k_1,k_2}\mathcal{A}_n(1,2,3,\ldots,n)+\mathcal{S}_{k_2,k_1+k_3}\mathcal{A}_n(1,3,2,\ldots,n) \nonumber \\
&+\ldots+\mathcal{S}_{k_2,k_1+k_3+\ldots+k_{n-1}}\mathcal{A}_n(1,3,4,\ldots,n-1,2,n) \label{Ima Plahte}
\end{align}
where we use the notation $\mathcal{S}_{k_i,k_j}\equiv \sin(2\pi \alpha'k_i\cdot k_j)$ and $\mathcal{C}_{k_i,
k_j}\equiv \cos(2\pi \alpha'k_i\cdot k_j)$. These are analytic relations between the amplitudes and so although
they are derived from the integral expression (\ref{open amp}) which only converges when all the momenta satisfy $2\alpha' k_i\cdot k_j>-1$. They will continue to hold for the analytic continuations of the amplitudes away from this restricted kinematic region.
In the limit $\alpha' \rightarrow 0$, The relation (\ref{real Plahte}) becomes  
\begin{align}
A_n(2,1,3,\ldots,n)=(-1)\sum_\sigma A_n(1,\sigma,n) \label{KK}
\end{align} 
where $\sigma \in OP(\{2\} \cup \{ 3,4,\ldots,n-1 \}) $ which is a set of ordered permutation preserving ordering within both sets, i.e. $\{2\}$ and $\{ 3,4,\ldots,n-1 \}$. The above equation expresses the Kleiss-Kujif relations in field theory \cite{KK-relations}.

Besides, when applying the same limit to the equation (\ref{Ima Plahte}), we obtain 
\begin{align}
0=&s_{12}A_n(1,2,3,\ldots,n)+(s_{12}+s_{23})A_n(1,3,2,\ldots,n) \nonumber \\
&+(s_{12}+s_{23}+s_{24})A_n(1,3,4,2,\ldots,n) \nonumber \\
&+\ldots+(s_{12}+s_{23}+\ldots+s_{1(n-1)})A_n(1,3,4,\ldots,n-1,2,n) \label{BCJ}
\end{align}
where $s_{ij}\equiv (k_i+k_j)^2=2k_i \cdot k_j$. This equation is exactly the BCJ relation \cite{BCJ}.

Generally, the Plahte identities expressed in (\ref{Plahte}) are valid for N-point amplitudes in both bosonic and supersymmetric string theory as transforming the integrand in the expression (\ref{open amp}) for the superstring theory from $z_{ij}^{2\alpha'k_i\cdot k_j}$ to $z_{ji}^{2\alpha'k_i\cdot k_j}$ encounters the same phase correction  (\ref{z-z}).


\section{KLT Relations} \label{appen KLT}
The KLT relations  were first derived by Kawai, Lewellen and Tye \cite{KLT} by factorizing a closed string scattering amplitude into a sum of products of color-ordered open string amplitudes. The general expression for $n$-point KLT relations was provided by \cite{MomentumKernel} as 
\begin{align}
&\mathscr{A}_n=(-i/4)^{n-3}\sum_\sigma\sum_{\gamma,\beta} \mathcal{S}_{\alpha'}[\gamma(\sigma(2,...,j-1))|\sigma(2,...,j-1)]_{k_1} \nonumber \\
&\times \mathcal{S}_{\alpha'}[\beta(\sigma(j,...,n-2))|\sigma(j,...,n-2)]_{k_{n-1}} \mathcal{A}_n(1,\sigma(2,\dots,n-2),n-1,n) \nonumber \\
&\times \widetilde{\mathcal{A}}_n(\gamma(\sigma(2,\dots,m-1)),1,n-1,\beta(\sigma(m,\dots,n-2)),n). \label{gen KLT}
\end{align}
where $\mathcal{A}_n$ and $\widetilde{\mathcal{A}}_n$ are left-moving and right-moving modes of open string amplitudes respectively. The object  $\mathcal{S}_{\alpha'}[\gamma|\sigma]_p$ is momentum kernel which maps products of open string amplitudes to closed string amplitude. The definition of the momentum kernel is given as
\begin{equation}
\mathcal{S}_{\alpha'}[i_1,\dots,i_k|j_1,\dots,j_k]_p\equiv \Big(\frac{1}{\pi \alpha'} \Big)^{k}\prod_{t=1}^k\sin\big(2\pi\alpha'\big(p\cdot k_{i_t}+\sum_{q>t}^k \Theta(i_t,i_q) k_{i_t}\cdot k_{i_q}\big)\big)
\end{equation}
with
\begin{equation*}
  \Theta(i_a,i_b)=\begin{cases}
    1, & \text{if $i_a$ appears after $i_b$ in the sequence $\{j_1,\dots,j_k$\}. }\\
    0, & \text{if $i_a$ appears before $i_b$ in the sequence $\{j_1,\dots,j_k$\}. }
  \end{cases}
\end{equation*}
Besides, $\mathcal{S}_{\alpha'}[\emptyset|\emptyset]_p=1$ for the empty set.

The expression (\ref{gen KLT}) comprises $(n-3)!\allowbreak(j-2)! (n-j-1)!$ terms and is independent of the value of $j$. The integer $j$ is arbitrary with $2\leq j \leq n-2$. Its value gives us choices to deform contours of integration for right-moving variables to be closed to the left or to be closed to the right around the branch points $x_1=0$ and $x_{n-1}=1$ respectively.

For $j=2$ and $j=n-1$ which is the case that all contours are totally closed to the right and left respectively, the expression consists of the greatest number of terms, namely ${(n-3)!(n-3)!}$. The choice made by the original KLT paper \cite{KLT} is to deform half of the contours to the left and the other half to the right, i.e. $j=\lceil n/2 \rceil$ corresponds to the minimum number of terms possible, $(n-3)!(\lceil n/2 \rceil-2)!(\lfloor n/2 \rfloor-1)!$ terms. 

\end{appendices}
\end{document}